\documentclass[letterpaper]{statsoc}
\usepackage{amsmath,amsfonts,amssymb,graphicx}
\usepackage[T1]{fontenc}
\usepackage{yfonts}
\newcommand{\Fig}[1]{Fig.~\ref{#1}}
\newcommand{\Sec}[1]{Section~\ref{#1}}
\def\Eqn#1{Equation~(\ref{#1})}
\def\Eqns#1#2{Equations~(\ref{#1}) and (\ref{#2})}

\def\card{{\rm card\,}}

\def\E{{\sf E}}
\def\Exp{\mbox{Exp}}
\def\N{{\mathcal N}}
\def\m{{\bf m}}
\def\h{{\bf h}}
\def\c{{\bf c}}

\title[Spatio-temporal structure in the onset of deposition]
{On building and fitting a spatio-temporal change-point model for settlement and growth at Bourewa, Fiji Islands}

\author[GK Nicholls {\it et al.}]{Geoff K. Nicholls}
\address{Department of Statistics,
         Oxford University,
         Oxford,
         UK.}
\coaddress{GK Nicholls, Department of Statistics, 1 South Parks Road, Oxford OX1 3TG, UK}
      \email{nicholls@stats.ox.ac.uk}
    \author{Patrick D. Nunn}
      \address{Department of Geography,
        The University of the South Pacific,
        Suva, Fiji}

\begin{document}

\begin{abstract}
The Bourewa beach site on the Rove Peninsula of Viti Levu is the earliest
known human settlement in the Fiji Islands. How did the settlement at
Bourewa develop in space and time? We have radiocarbon dates
on sixty specimens, found in association with evidence for human presence,
taken from pits across the site.
Owing to the lack of diagnostic stratigraphy, there is no direct archaeological evidence for
distinct phases of occupation through the period of interest.
We give a spatio-temporal analysis of settlement at Bourewa in which the deposition rate for
dated specimens plays an important role.
Spatio-temporal
mapping of radiocarbon date intensity is confounded by
uneven post-depositional thinning.
We assume that the confounding
processes act in such a way that the absence of dates remains informative of zero rate for the
original deposition process.
We model and fit the onset-field, that is, we estimate for each
location across the site the time at which deposition of datable specimens began. The temporal
process generating our spatial onset-field is a model of the original settlement dynamics.
\end{abstract}

\maketitle

\section{Introduction}

\begin{figure}[htb]
  \hspace{-0.5in}\includegraphics[width=7in]{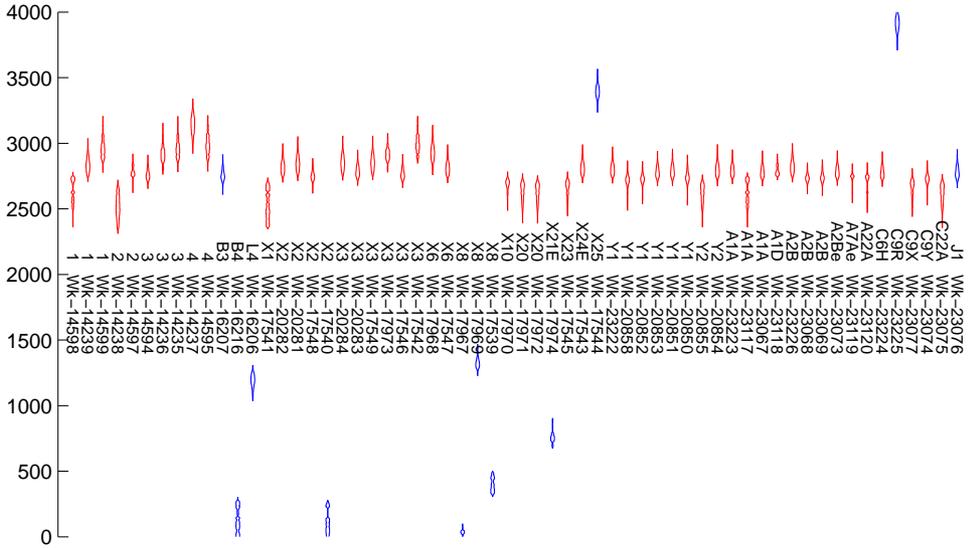}\\[-0.25in]
  \caption{All Bourewa data: (y-axis) years BP, (x-axis) specimen index, likelihoods (blue) data omitted from study, (red) data retained.
  Two early dates were ajudged insecurely linked to evidence for human presence, and the later
  dates belong to a qualitatively different phase of activity at the site.
  Text (top row) is pit-name and (bottom row) date identifier.}\label{fig:bourewadata}
\end{figure}

\begin{figure}[htb]
  \[\includegraphics[width=4in]{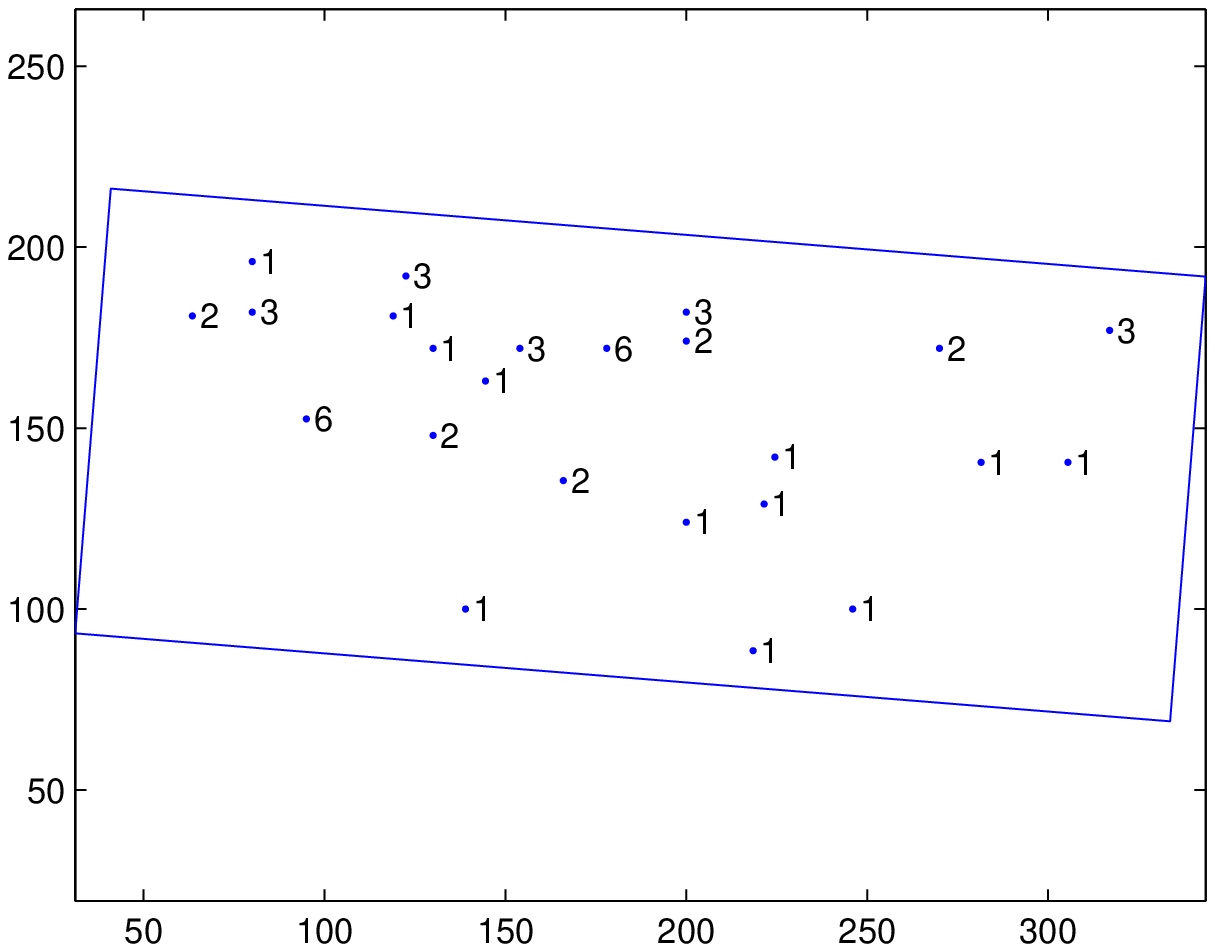}\]
  \caption{The scatter of pit locations, and the number of dated specimens in
    each pit which were retained in the analysis. The box marks the edges of the
    onset-field we will model and has been aligned to the data. The sea is adjacent to
    the upper edge of the box, with the short box axis pointing towards the sea,
    and the long axis pointing along the beach. Axes give excavation coordinates.}\label{fig:pitnumscat}
\end{figure}

Bourewa beach is an archaeological site on the southwest coast of Viti Levu, the largest
island in the Fiji Islands Group.
It was identified in \cite{nunn04} as a site of likely early settlement by surface finds of distinctive
early Lapita-era ceramics.
In the course of the excavation, which ended in February 2009, in excess of one hundred pits,
of varying sizes, were dug.
Material was selected from some of these pits for radiocarbon dating.
These data, and their observation model,
are described in \Sec{sec:data}. The likelihoods for the unknown ages of the
specimens are plotted in \Fig{fig:bourewadata}.
The number of dates measured in
any given pit varies from zero to six. The relative pit locations and the number
of dated specimens in each pit are shown in \Fig{fig:pitnumscat}.
Criteria for selection for dating included secure
association with cultural remains, for example, human burials and diagnostic forms of decorated pottery.
The pits were excavated over several years, and dated incrementally, under funding constraints
which varied through time. There is a need to date across the site in three spatial dimensions,
so material selected for dating is to some extent deliberately spread out.

There is a conjecture, described in \cite{nunn07,nunn09}, that
a small initial settlement in the centre of the beach grew in size
to eventually cover the studied site area. We formalise this conjecture,
and compute the posterior probability that it is true, along with other measures
of support. We make a spatio-temporal map of the date at which the deposition
of human-related material began at each point on the site.
For each spatial location, this ``onset-field'' map gives the time
at which the deposition rate changes from zero to some positive value.
The inference is conditioned to ensure that the pit locations and the distribution
of dated specimens across pits are not by themselves informative of variations
in the time-span of human activity from pit to pit.

We begin our modeling of the data with a description, in \Sec{sec:phasemodel}, of a
family of prior models for the joint distribution of
specimen ages. These temporal models, developed in
\cite{naylor88,zeidler98,nicholls01} and implemented in the OxCal software
described in \cite{ramsey01}, are in widespread use.
The dated-specimen deposition process is a Cox process in time.
Specimen ages
are the times for events in a Poisson point process, which has a rate which may vary from pit to pit,
but is constant in time between change points. The site-wide dated-specimen deposition process is
the sum of independent Poisson processes in each pit (the specimen are all small pieces of
charcoal and shell and would have been waste material at the time).
Change points
are the boundaries of phases of deposition. In this class of models, the number of
change points is known from separate evidence. Phase boundaries mark site-wide
changes in the human activities generating dateable specimens and may be indicated
by strata.
The phase boundaries are modeled as events in a second constant rate Poisson point process.
The rate parameters (for specimen ages, and phase boundary ages)
are not usually physically interesting, or easy to model, as the ages
of dated specimens are the ages of events
in an original deposition process thinned by arbitrary specimen selection at excavation,
and the uneven action of decay.
It is not necessary to
estimate these rates if we can condition the analysis on the number of phases,
and the number of dates in each phase.
This is possible, for example, if phases are strata and a fixed
number of specimens are taken temporally at random within known strata.

\begin{figure}[htb]
  \includegraphics[width=6in]{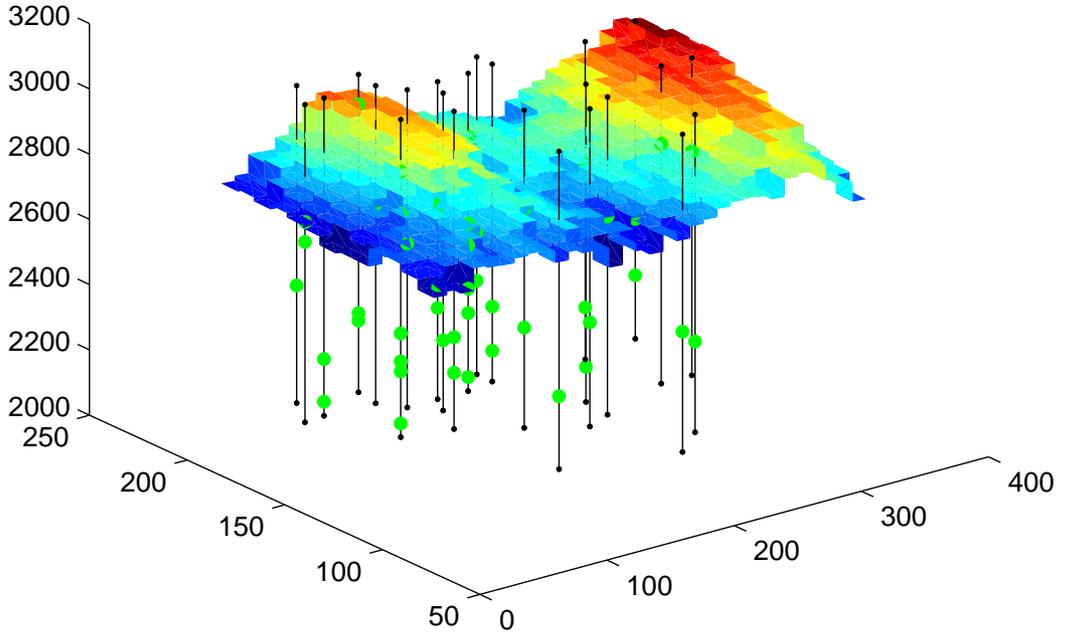}\\
  \vspace*{-0.25in}
  \caption{A realization of the prior of \protect\Sec{sec:onsetfieldbourewa}
  for specimen ages $\theta$ (green dots),
  phase boundaries (black dots, for a single phase, $M=1$), onset-field $\phi$
  (piecewise constant surface of $32\times 13$ cells) for $\alpha$,$\beta_1$ and $\beta_2$
  distributed with $A=10$ and $B=1$ in \protect\Eqn{eq:alphabetaprior} and limits
  $L=2000$, $U=3500$ years BP. (z-axis) years BP,
  (x,y-axes) 0.25cm units.}
  \label{fig:priorrealizationNPYp0N11N60}
\end{figure}

In our spatio-temporal model, described in \Sec{sec:onsetfield},
an onset-time field replaces the earliest phase boundary of the
temporal phase model of \Sec{sec:phasemodel}. This
allows the start-time for the deposition of dateable material
to be a function of position. The phase and dated-specimen
processes are otherwise unchanged.
The onset-field process is
a two-parameter spatio-temporal process on a rectangular lattice, and is related to
the family of models described in \cite{Richardson73}.
One parameter of the onset-field process controls the rate at which an unoccupied cell is occupied
by ``immigration from off-site'' and the other the rate for cell occupation from neighboring
occupied cells. For suitable parameter
choices, the field can vary from a simple space-time cone, developing like a solution to the
non-linear Fisher equation in a wave-of-advance, to a randomly corrugated surface.
A prior sample is shown in \Fig{fig:priorrealizationNPYp0N11N60}.
Further aspects of the model are illustrated in the supplement.
We fit this model in \Sec{sec:firstfit}, allowing just one phase, since there
is no given phase structure for the Bourewa data.
In this model, dated specimens are generated in a single interval with a spatially varying onset-time
and a finishing time which is the same everywhere. The dated-specimen deposition process has a rate
which is assumed constant in time when it is non-zero, but may have arbitrary spatial structure.
Because we condition the analysis on the number of dates in each pit and on the number in each
phase (there is only one phase), there are no deposition rate parameters left to estimate.

In the remainder of the paper we fit extensions to the single-phase onset field model.
However, the model in \Sec{sec:firstfit} has a good balance of realism and simplicity,
and the further development may be read as model mis-specification testing.
Is there really just a single phase, and if not, what temporal variation is present?
Although there is no evidence from stratigraphy for phase structure for the Bourewa data,
it is possible that
some phase structure is present. This possibility is implicit in
the analyses of \cite{nunn07,nunn09} and is a second question (besides the
settlement hypothesis) of archaeological interest.

The third model, which we describe in \Sec{sec:phasestructure},
is a non-parametric extension of the basic phase model of \Sec{sec:phasemodel},
with no spatial structure. The number of change points
in the deposition rate between the beginning and end of deposition is unknown.
Since the phase structure is unknown,
we can no longer condition on the assignment of specimens to phases. We need therefore to model the ratio
of deposition rates between phases, in order to weight the assignment of dates to phases.
Significant variation in the accumulation
rate of specimen dates is evidence for phase structure.
However, 
although evidence for two or more phases
 is evidence for rate variation, the phases we reconstruct need not
 be culturally significant.
We fit the phase structure model of \Sec{sec:phasestructure}
to the Bourewa data in \Sec{sec:fitphasestructurebourewa} and get
results which are misleading in just this way.
We therefore test this third model in \Sec{sec:fitphasestructurestud}
with an application to a
second data set, the Stud Creek radiocarbon data, published by \cite{holdaway02}.
For these data there is a hypothesis
that the original deposition was interrupted for some time. We reject the
single-phase model in favor of a three phase model.
We are at this point fitting a change-point process of unknown piece-wise constant rate,
with no spatial component. \cite{green95} uses a temporal change-point analysis of
coal-mining fatalities data to illustrate reversible-jump MCMC.
Our event times are observed with the radiocarbon uncertainty, but the analysis is,
at this point, otherwise the same.

We return in \Sec{sec:fullmodel} to the Bourewa data to fit a model with
the randomly variable temporal phase-structure of \Sec{sec:phasestructure} and the spatio-temporal onset-field
of \Sec{sec:onsetfield}.
In this fourth and final model, we start deposition with a spatially varying onset field
and follow this with an unknown number of phases. We condition on the assignment of dates to pits,
but not the assignment of dates to phases. The assignment
of a specimen to a phase depends on the ratios of deposition rates between phases,
as in \Sec{sec:phasestructure}.
We assume that these ratios do not vary from pit to pit.
This reduces the number of unknown rate parameters to a small number,
one less than the (unknown) number of phases.
The assumption is good if, for example, it was true for the original deposition,
and the uneven thinning due to specimen selection
and decay is separable, so that the thinning probability is the product
of one function of space and another of time.

\section{Related literature}

Discusion of related literature on the Richardson process itself is given in \Sec{sec:richardson}.

\subsection{Methodology}

\cite{majumdara05} give a Bayesian statistical framework for the analysis
of spatio-temporal data with a single change point.
They note that the framework is easily generalised, and illustrate it
by fitting a Gaussian process $Y(s,t)$
$s\in \Re^2$, $t\in \Re$ with a change point at time $t=t_0$.
The process changes from $\mu_t+U(x,t)+W(x,t)+\varepsilon(x,t)$
at $t\le t_0$ to $\mu_t+U(x,t)+W(x,t)+\varepsilon(x,t)$ at $t>t_0$. Here $\varepsilon(x,t)$
is a field of independent normal random variables,
and $U$, $V$ and $W$ are Gaussian processes with separable translation-invariant
covariance functions. These data are observed at $N$ points in space at each of $M$
points in time. In their successful simulation study they fit fourteen parameters
to data: the scalar means and variances of $\varepsilon$ either side of $t_0$ (four), the space and time
scales of the exponential class correlation functions for $U,V$ and $W$ (six)
and their three variance-scaling
parameters, plus $t_0$ itself. Our own fitting assumes complete spatial independence between
specimen-deposition events at different pits, conditional on the unknown deposition rate.
On the other hand, we do not know the deposition rates, or the
number of change points, and the onset field is a new kind of change-point,
since it is a change-field which spreads gradually over the region,
marking the transition from zero to non-zero deposition rate. \cite{ibanez07}, who wish
to model change-points which spread in this way, have suggested a modification
of \cite{majumdara05}, though there is to date no fitting.

We found few applications of Bayesian or likelihood-based inference for
spatial-temporal growth processes. \cite{zhu08} fits a complex, and relatively realistic model
to beetle presence and count data and tree-mortality data for two beetle
species and the tree-health indicator simultaneously. The two species compete
and spread on a random graph with trees at nodes. Time is discrete corresponding to year.
The likelihood for the process is given as a Markov Random Field up to an unknown normalising constant.
The process (beetle counts, tree state) is observed directly.
Our onset-field is similar: the process is spread $via$ neighbors, and immigration.
The probability density for the onset-field process here is
given as an MRF also, though our simpler process is normalisable.
However, our onset-process is observed indirectly, through the events which follow it,
and it is this aspect, its role as a change point, which distinguishes the applications.

\cite{moller08} fit
data for a point process $\mathbf{X}_t$ of bush-fire locations and discrete times
allowing a spatially structured intensity of the form $\lambda(x,t)=\lambda_1(x)\lambda_2(t)S(x,t)$
with ${\sf E}(S(x, t)) = 1$ and  $(x, t) \in \Re^2 \times \mathbb{Z}$.
The components
$\lambda_1$ and $\lambda_2$ are given in terms of linear predictors from covariates for
spatial and temporal structure respectively and \cite{moller08} take a shot-noise process
for the residual process $S(x,t)$. \cite{diggle05} set this
framework up to model spatial-temporal disease data, with a unit mean log-Gaussian
residual process.
Because the likelihood is intractable, \cite{moller08} develop
alternative estimation procedures, including Lindsay's composite likelihood.
The aim in \cite{diggle05} and \cite{moller08} is to model the bulk point process
in 2+1D. In contrast, our data collapse the spatial locations of dated specimens in a pit
(which has some extension) onto a single point, and
events in each pit are modeled using a process which is marginally a 1D Cox process.
Separable intensities play a role in all the spatial-temporal statistical analyses we have cited,
including our own.

\subsection{Application}

Where there is an interest in recovering unknown phase structure without spatial structure,
many authors sum the likelihood functions for individual specimen ages to get
a single function of age. \cite{holdaway08}
is a rare example in which this procedure may be justified. Under certain conditions, the
summed likelihood is an estimate of the dated-specimen deposition-rate function.
\cite{bayliss07} warn against the common practice
of treating this as a proxy for population density.

Spatial maps based on the
distribution of radiocarbon ages are not usually model based, but simply a sequence of
scatter plots of find locations windowed by age, as \cite{graham96}.
One dimensional projections of 2D data, such as the
compilation of early European-Neolithic dates by \cite{ammerman71}, who projected on
distance-from-Jericho, are commonly regressed. \cite{hazelwood04} give a recent
overview of the same subject, and make a comparison with the settlement of North America,
in two spatial and one temporal dimension. These authors fit
a Fisher wave-of-advance model using a linear regression to get the wave speed,
and guess plausible values for other model parameters based on prior knowledge of
generic human demography.
In the latest work in this spirit, \cite{davison09} map the Neolithic settlement of Europe, modeling
non-interacting expansion from
two sources, using a Fisher wave-of-advance model with
parameters for growth rate, carrying capacity, spatially varying 2D advection and
a scalar diffusivity.
They pool or otherwise reduce the earliest dates at a given location to
form a central estimate for the local arrival
of the Neolithic peoples. They fix model parameters using a hybrid scheme.
Initial conditions, such as the starting locations and times
of the two expansions, are obtained by minimizing the root mean square difference between the reduced date at each
location and the arrival time of the modeled population wavefront at that location.
This is nonlinear regression.
Parameters of the advection-diffusion are fixed using prior knowledge of
generic human demography, as in \cite{hazelwood04}.
Note that, although these authors
do not make likelihood-based inference, they are modeling an onset-field.

\cite{blackwell03} treat a collection of radiocarbon dates from a number
of sites excavated under different conditions at different times over northern Europe.
For such spatially sparse data it is not useful to couple the onset event at each site
with a smooth spatial field. They group the data into regions, and fit an independent
temporal model for each region.
Since the analyses  are not spatially coupled, \cite{blackwell03} page 238
describe their own analysis as ``not truly spatiotemporal in nature''.
In contrast, the data we treat comes from a single intensively studied site.
Dated pits are dense in the modeled region. These properties of the Bourewa data
justify some spatial smoothing of the field of onset times.

\section{Data, and observation model}
\label{sec:data}


The data are $K=60$ uncalibrated radiocarbon ages $y_i,i=1,2,\dots K$, with associated laboratory standard
errors $\sigma_{i}$, grouped in $H=32$ pits. Some of these data were dropped
as discussed in the final paragraph below. For pit number $h=1,2,\dots,H$, $x_h=(x_{1,h},x_{2,h})$ gives
pit coordinates.
Let $\h(i)$ map data index to pit index.
The location $x_{\h(i)}=(x_{1,\h(i)},x_{2,\h(i)})$ of the $i'th$ dated specimen
is given as a point-location for the pit in which it was found, although pits vary in size from 0.25
to 2 meters on each side. Some pits are dug but not dated. Dated pits are distributed over a region approximately 200 by 50 meters,
with the majority of pits in a cluster some 50 meters square.

Denote by $\Theta_i$ the unknown true age of the $i$'th dated specimen,
measured in calendar years before 1950 (Before the Present, BP). Observations of specimen ages
are distorted by a non-linear, empirically determined, calibration function $\mu$,
with associated error function $\sigma$.
In the standard observation model of \cite{stuvier77,buck92},
\[Y_i=\mu(\Theta_i)+\epsilon_i\]
with $\epsilon_i\sim N(0,\sigma_i^2+\sigma(\Theta_i)^2)$ independent random variables.
We are omitting some
straightforward details: marine and terrestrial material
are in fact calibrated using different calibration functions 
and radiocarbon ages measured on local seashell
are subject to the small constant marine-reservoir offset given in \cite{nunn07}, which
we treat as discussed in \cite{jones01}. In our software we use the 2004 calibration functions
of \cite{mccormac04} and \cite{hughen04} interpolated from their calibrations at 5 year intervals down to one year
intervals, and then approximate $\Theta_i$ as an integer year. In our discussion here
$Y$ and $\Theta$ are continuous.

The normal likelihood for parameters $\Theta=\theta$ given data $Y=y$ is
\[\ell(\theta;y)\propto \prod_{i=1}^K (\sigma_i^2+\sigma(\theta_i)^2)^{-1/2}\exp\left(-\frac{(\mu(\theta_i)-y_i)^2}{2(\sigma_i^2+\sigma(\theta_i)^2)}\right)\]
with $\mu$ and $\sigma$ functions of $\theta$ available from a look-up table.
The likelihoods of the Bourewa specimen ages are graphed in \Fig{fig:bourewadata}.
Certain specimens (blue points) were dropped from the analysis as either insecurely
linked to evidence of human presence, or associated with activity outside the period of interest.
Some pits containing a single omitted specimen were thereby removed from the analysis. Data from two pits lying far from the central area were dropped as the spatial
extrapolation to those isolated pits was deemed too great, leaving the $H=24$ pits
and $K=49$ radiocarbon dates represented in \Fig{fig:pitnumscat}.

\section{Conditioning on phase structure}
\label{sec:phasemodel}

The posterior distribution specified in this
section can be simulated using existing OxCal software. However,
the notation of this section is needed below,
and the numerical results will be of interest because, although
obtained from standard models in the literature, they conflict with the results we get from
the onset-field model.

We begin with a prior model of the process generating dates in a single pit.
Dates within a pit are grouped into
phases, corresponding to some form of strata. The ages $\Theta_i,\Theta_j$ of specimen
from distinct strata are known {\it a priori} to be ordered, so that $\Theta_j>\Theta_i$
if specimen $j$ is from a deeper stratum than that of specimen $i$. Denote by $\Psi_m, m=0,\dots,M$
the unknown true age of the $m$'th phase boundary, with $\Psi_m<\Psi_{m+1}$, so that numbering is from the
surface to the bottom of the pit, and the $m$'th phase is the age interval $[\Psi_{m-1},\Psi_{m}]$.
Let $\m(i)$ give the known phase ($ie$, stratum) of
the $i$'th dated specimen. Phases may be empty, with no specimen taken from the corresponding stratum,
as happens if for example there is a hiatus in deposition, or possible erosion.
Let $L$ and $U$ be fixed lower and upper bounds on specimen and phase
boundary ages, available as prior knowledge. We use $L=2000$ and $U=3500$ years BP
except in \Sec{sec:fitphasestructurestud}. The parameter state space for analysis
of a single pit is
\[\Omega=\{(\psi,\theta):
L<\psi_0,\ \psi_{m-1}<\psi_{m}\ m=1,..,M,\ \psi_M<U,\ \psi_{\m(i)-1}<\theta_i<\psi_{\m(i)}\ i=1,..,K\}.\]

We follow \cite{nicholls01}, who get a prior density for $(\Psi,\Theta)$ by
modeling the deposition process generating those parameters. The phase boundaries
$\Psi_1,\dots,\Psi_{M-1}$ are the events of a Poisson point process $\Lambda_\Psi$
of constant rate $\lambda_\psi$ which starts at the onset-time $\Psi_M$ and stops at
the ending-time $\Psi_0$.
\cite{nicholls01} prior density for onset and ending times $\Psi_M=\psi_M$ and $\Psi_0=\psi_0$
is $p_{0,M}(\psi_0,\psi_M)=(U-L-(\psi_M-\psi_0))^{-1}$
and this leads to a joint prior density $p(\psi|M)$ at $\Psi=\psi$ equal to
\begin{equation}\label{eq:psigivenM}
p(\psi|M)=\frac{1}{(U-L-(\psi_M-\psi_0))}\frac{1}{(\psi_M-\psi_0)^{M-1}}.
\end{equation}
This density has a uniform marginal distribution for the span statistic, $\Psi_M-\Psi_0\sim U(0,U-L)$.
This is desirable when the span is of particular interest, as is often the case
(though \cite{naylor88} make $p(\psi|M)=1$ favoring greater spans over lesser).
The specimen ages $\Theta_1,\dots,\Theta_K$ are
the events of a Poisson point process of piece-wise constant rate $\lambda_\theta(t)=\lambda_{\theta,m}$
for $\psi_{m-1}<t<\psi_m$. The rates $\lambda_{\theta,m}$ are not of interest,
since the archaeologist imposes an uneven thinning on specimen, in selecting specimen for dating,
on top of possible uneven thinning by in situ decay. However, since the analysis is
conducted conditional on the number $M$ of phases, and conditional on specimen phases $\m$,
the conditional density for $\Theta|\Psi$ is
\begin{equation}\label{eq:thetagivenpsim}
p(\theta|\psi,\m)=\prod_{i=1}^K (\psi_{\m(i)-1}-\psi_{\m(i)})^{-1}.
\end{equation}
The posterior distribution is
\begin{equation}\label{eq:standardpost}
p(\psi,\theta|y,\m,M)=\ell(\theta;y)p(\psi,\theta|M,\m),
\end{equation}
with
\[
p(\psi,\theta|M,\m)=p(\theta|\psi,\m)p(\psi|M)
\]
given by \Eqns{eq:psigivenM}{eq:thetagivenpsim}.

\begin{figure}[htb]
  \hspace*{-0.5in}\includegraphics[width=6.5in]{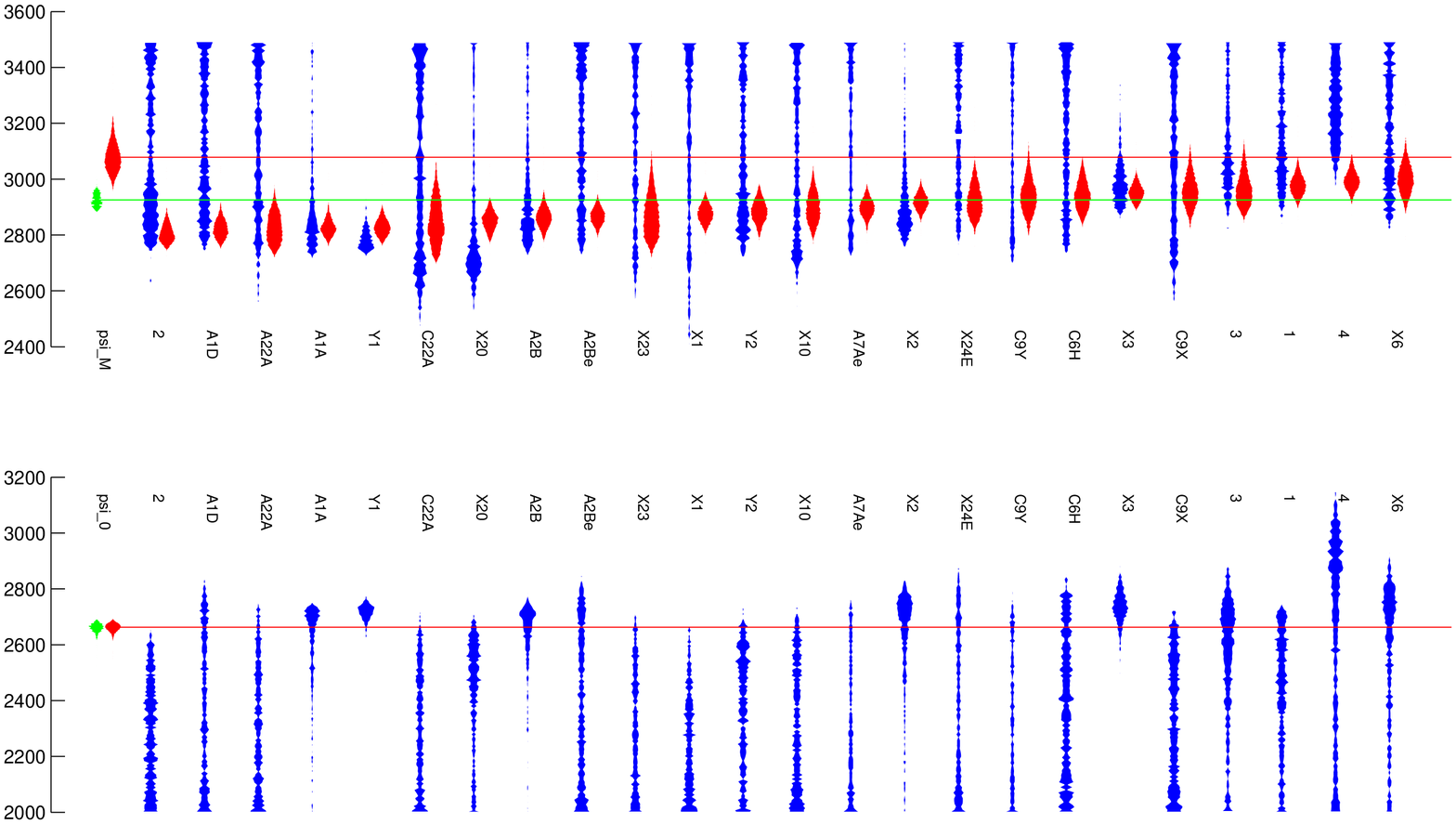}\\
  \vspace*{-0.25in}
  \caption{(top) posterior distributions of onset-times
  ($\psi_M$ or $\phi_{\c(h)}$, y-axis, years BP)
  by pit (pit-names, x-axis, sorted by median onset). (bottom) posterior distributions
  of ending times $\psi_0$. Leftmost histograms labeled $\tt psi\_M$ are (red) $\psi_M$ for the onset
  field analysis of \protect\Sec{sec:firstfit} and (green) $\psi_M,\psi_0$ for all dates in single phase, as \protect\Sec{sec:phasemodel}.
  The remaining columns are (red) onset-field interpolations $\phi_{\c(h)}$ and (blue) $\psi_M,\psi_0$ from
  an independent single phase analysis (as \protect\Sec{sec:phasemodel}) of the data from the pit for that column.}
  \label{fig:pitages}
\end{figure}

In this kind of analysis, a site is treated as in effect a single pit.
Fitting this model to the Bourewa data, for a single phase, $M=1$, so $\m(i)=1$
for each $i=1,2,\dots,K$
using the MCMC algorithm given in \cite{nicholls01} to sample the posterior distribution
of \Eqn{eq:standardpost}, we get the marginal posterior distributions
for onset age $\psi_M$ and ending age $\psi_0$ shown in green in the leftmost
column of \Fig{fig:pitages}.
As part of a later model mispecification analysis, we fit the (single phase)
model of this section to each pit independently, including pits with just one or two dated specimens
(for which the posterior distribution of $\psi_1$ and $\psi_0$ is dominated by the choice
of the upper and lower limits, $U$ and $L$). These distributions are displayed in blue
and labeled by pit name in \Fig{fig:pitages}.

\section{A spatial-temporal onset-field model}
\label{sec:onsetfield}

The whole process is started at the first-onset time $\psi_M$. This is ``first settlement''.
The onset field begins to evolve, as the area in use is enlarged
by spreading out, and by new arrivals. New phase events at times $\psi_{M-1},\dots\psi_1$ are
step-changes in the deposition rate across the then-occupied site,
corresponding to changes in the way the then-occupied site is used. The
deposition process terminates at the ending event, at time $\psi_0$.

\subsection{The onset-field process}
\label{sec:onsetalgsec}

Let $\phi_c,\ c=1,2,\dots,C$ be a field $\phi_c\le \psi_M$,
defined on a rectangular $C_1\times C_2$ lattice of $C=C_1C_2$ square cells.
Let $\N(c)$ give the set of neighbors of the $c$'th cell. Corner, side and interior
cells have two, three and four nearest neighbors respectively. Let $\N^+(c)=\N(c)\cup\{c\}$.
Let $\c(x)$ map points $x$ in the pit coordinate system to the index of the overlaying lattice cell.
A modeled pit has a point-like location, though it may lie cross cells.
We report results for a $13\times 32$ lattice of cells each
about 2.375 meters square. 

The variable $\phi_{c}$ gives the onset time for deposition at points
$x$ in cell $c=\c(x)$.
The onset time $\phi_{c}$ in cell $c$ is the time of first occupation
for cell $c$. There is one arrival
in each cell and $C$ arrivals in the course of the process.
Let $r_{c,n}$ give the arrival rate
in cell $c$ in the time interval preceeding the $n$'th arrival in the process,
and let $R_n=\sum_c r_{c,n}$ give the total arrival rate in the $n$'th interval.
Let $L<\psi_0<\psi_1<\dots\psi_M<U$ and for $c=1,2,\dots,C$ and $c'\in \N(c)$
let positive constants $\alpha_c$ (the rate for immigration to cell $c$) and $\beta_{c,c'}$
(the rate for occupation of cell $c'$ from occupied cell $c$, with $\beta_{c,c'}=\beta_{c',c}$)
be given. We call the process parameterised by $\beta$ `migration'.

The onset-field $\phi$ is a realization of the following immigration-migration process:
cell $c$ is occupied by immigration at rate $\alpha_c$; when cell $c$ becomes occupied,
the arrival rate for occupation at each of its neighbors $c'\in \N(c)$ is {\it increased} by $\beta_{c,c'}$.
A cell once occupied stays occupied.
When we fit data, we will condition on the first onset-field event occurring
at $\psi_M$ years BP; the algorithm below is for the unconditional case.

\begin{minipage}[c]{5in}
\begin{description}
\item[]
\item[Step 1] {\it Initialize field of rates}
\begin{description}
  \item[Step 1.1]  $t_0\leftarrow \psi_M$, $r_{c,1}\leftarrow \alpha_c$ for each $c=1,2,\dots,C$
    and $R_1=\sum_c r_{c,1}$.
    \item[Step 1.2] Set $n\leftarrow 1$.
        \end{description}
  \item[Step 2] {\it Simulate arrivals at {\rm Exp$(R_n)$} intervals:}\\[0.05in]
    while $R_{n}>0$\\[-0.21in]
    \begin{description}
    \item[Step 2.1] simulate $\delta_n\sim \Exp(R_n)$ and
        $c_n\sim (r_{1,n}/R_n,r_{2,n}/R_n,\dots,r_{C,n}/R_n)$;
    \item[Step 2.2] Set $t_n\leftarrow t_{n-1}-\delta_{n}$ and $\phi_{c_n}\leftarrow t_n$.
        For each $c=1,2,\dots,C$, set $r_{c,n+1}\leftarrow r_{c,n}$ and then overwrite
        $r_{c_n,n+1}=0$ and $r_{c,n+1}\leftarrow r_{c,n+1}+\beta_{c_n,c}$
        for each $c\in \N(c_n)\setminus\{c_1,c_2,\dots,c_{n-1}\}$.
\item[Step 2.3] set $n\leftarrow n+1$.
\item[]
    \end{description}
\end{description}
\end{minipage}

We orient the lattice along the beach, and
allow the rates for migration along and at right angles to the
long axis of the beach to be unequal. We fit fields for $\alpha_c=\alpha$ and
$\beta_{c,c'}=\beta_1$ for $c\rightarrow c'$ along the beach and $\beta_{c,c'}=\beta_2$
for $c'$ closer to, or further from, the sea.
If $\alpha$ is small, and $\beta_1=\beta_2=\beta$ is large, typical onset fields show a spread from
a single centre. A 2+1D plot is cone-like. As $\alpha$ increases at fixed $\beta$,
the number of centres-of-expansion increases, corresponding to isolated settlements which grow and merge.

The onset-field model is convenient for fitting.
The joint density $p(\phi|\alpha,\beta,\psi)$ of $\phi$ is
\[
p(\phi|\alpha,\beta,\psi)=\prod_{n=1}^C r_{c_n,n}\exp(-R_n\delta_n),
\]
from the algorithm. 
Now \[\sum_n R_n\delta_n=\sum_{c=1}^C\left[\alpha_c(\psi_M-\phi_c)+
                       \frac{1}{2}\sum_{c'\in\N(c)} \beta_{c,c'}|\phi_c-\phi_{c'}|\right].\]
                       Each cell $c$ contributes to $\sum_n R_n\delta_n$ a term
                       $\alpha_c(\phi_c-\psi_M)$ and for each
                       neighbor $c'$,
                       a term $\beta_{c',c}(\phi_{c'}-\phi_{c})\mathbb{I}_{\phi_c<\phi_{c'}}$.
                       Taking the two such terms for each edge gives
                       $\beta_{c,c'}|\phi_c-\phi_{c'}|$.
Let $\rho(\phi_{\N^+(c)})$ be the rate function of $\phi_c$,
\begin{equation}\label{eq:rho}
    \rho(\phi_{\N^+(c)},c)=\alpha_c+\sum_{c'\in \N(c)}\beta_{c,c'}\mathbb{I}_{\phi_{c'}>\phi_c},
\end{equation}
so that $\rho(\phi_{\N^+(c_n)},c_n)=r_{c_n,n}$.
This is the arrival rate for occupation of cell $c$ in the time interval preceding $\phi_c$,
and $\phi_c$ is an age BP, so that interval is ages greater than $\phi_c$.
Assembling these terms gives
\begin{equation}\label{eq:phi}
    p(\phi|\alpha,\beta,\psi)=
    \prod_{c=1}^C\exp\left(\log(\rho(\phi_{\N^+(c)}),c)-\alpha_c(\psi_M-\phi_c)-
                            \frac{1}{2}\sum_{c'\in\N(c)} \beta_{c,c'}|\phi_c-\phi_{c'}|
                            \right).
\end{equation}
The right hand side is normalized over $\phi\in (-\infty,\psi_M]^C$ so inference for
$\alpha$ and $\beta$ is not obstructed by an intractable normalization.
Also, $\Phi$ is a Markov Random Field (MRF)
with cliques $\{c\},\{c,c'\}_{c'\in \N(c)}$ and $\N^+(c)$.



%

\subsection{The Richardson growth model, and related models}
\label{sec:richardson}
The onset-field process is a special case of the contact process
with immigration. Cells of a contact process, occupied by the
immigration and migration processes above, are abandoned at {\it per capita} rate $\gamma$; a
binary field $\xi_c(\tau),\ c=1,2,\dots,C,\ \tau\in [0,\infty)$ indicates
if a cell is occupied at time $\tau$. The process is often studied without immigration,
conditioned on one or more arrivals, or seeds, at time $\tau=0$.
The equilibrium statistics of $\xi$ are studied, as functions of $\beta$ and $\gamma$.
\cite{durrett94} review the process in the context of applications in ecology.
A contact process in which a single cluster of occupied cells grows from a single seed,
with no immigration or abandonment ($\alpha=\gamma=0$), is a Richardson cluster-growth model,
so the onset-field model above is a Richardson model with multiple clusters, $ie$, with immigration.
\cite{Richardson73} showed that the cluster grows linearly
and tends in shape towards some fixed shape. \cite{durrett81} extend
this, giving further results for the boundary shape. If the process is isotropic,
this limiting shape is a circle.
\cite{hammersley77} and \cite{durrett81} are sceptical. However, the
limiting shape is not known at present. The simulation experiments reported in the Supplement
show that it is at least very nearly isotropic. This is important as we would like
the process at $\beta_1=\beta_2=\beta$ to be isotropic, for clarity of interpretation.

\cite{lee99} uses a Richardson process with a Poisson number of
randomly located seeds to model fields of aluminium grains
as random lattice segmentations. The
cluster seeds all 'arrive' at the start time.
The process realizes a
random segmentation of the lattice into regions
labeled by the index of their seed. \cite{lee99} conjectures that,
for his initialisation, the large scale structure of the $\alpha=0$, $\gamma=0$ process is isotropic,
and for fixed seeds the region borders converge to the edges of a Voronoi tesselation
as the cells are subdivided.

Because our seeds arrive at different times,
the onset-field process generates a segmentation which looks something like a
Johnson-Mehl tesselation at large scale.
Johnson-Mehl tesselations, described for example in \cite{moller99},
are random tesselations of the plane continuum.
Seeds arrive in $\mathbb{R}^2\times [0,\infty)$ according to a Possion process in two
space and one time dimension. Each seed captures a circular domain which grows
with a fixed speed until it meets the boundary of another domain. Seeds falling into existing
domains are deleted. This continuum version of the process we are using,
with deterministic region growth, would be a natural off-lattice model,
which we have not considered.

See the supplement for further discussion of the stationarity and isotropy of the
onset-field process, and a relation to traveling wave models.

\subsection{The onset-field as a prior probability density for the Bourewa data}
\label{sec:onsetfieldbourewa}
The full prior density for the inference in the next section is
\begin{equation}\label{eq:fullprior}
p(\alpha,\beta,\theta,\phi,\psi|M,\m)=p(\theta|\phi,\psi,\m)p(\phi|\alpha,\psi,\beta,\max(\phi)=\psi_M)
                                      p(\psi|M)p(\alpha,\beta).
\end{equation}
Prior density $p(\psi|M)$ is given \Eqn{eq:psigivenM}.
We condition the onset-field on $\max(\phi)=\psi_M$; in terms of model elements,
this means that the beginning of the earliest phase coincides with first settlement,
and spreading occupation.
The modification is
\[
p(\phi|\alpha,\beta,\psi,\max(\phi)=\psi_M)=\alpha^{-1}p(\phi|\alpha,\beta,\psi)
\]
for $p(\phi|\alpha,\beta,\psi)$ given in \Eqn{eq:phi}, and
\[\phi\in \{\phi: \phi\in (-\infty,\psi_M]^C,\max(\phi)=\psi_M\}.\]
Although we now drop $\max(\phi)=\psi_M$ in our notation, all the distributions
below which involve $\phi$ are conditioned in this way.
The onset field fixes the local change
point from zero to non-zero rate for dated-specimen deposition. This imposes
\[\psi_{\m(i)-1}<\theta_i<\min(\phi_{\c(x_{\h(i)})},\psi_{\m(i)}).\]
The bound on the right hand side is the change-point to a non-zero deposition rate,
so we should replace the density for $\Theta|(\Psi,\m)$ given in
\Eqn{eq:thetagivenpsim} with
\begin{equation}\label{eq:thetagivenphipsim}
p(\theta|\phi,\psi,\m)=\prod_{i=1}^K \frac{1}{\min(\phi_{\c(x_{\h(i)})},\psi_{\m(i)})-\psi_{\m(i)-1}}.
\end{equation}
The onset field can be thought of as a spatio-temporal mask over temporal
phase structure which would otherwise apply at all locations.

Our prior for $\alpha$ and $\beta$ is subjective. We expect just a handful of
pure immigration events in the interval $[L,U]$, so we need
$\E(\alpha)\gtrsim 1/C(U-L)$. The ``speed'' of the
expansion is controlled by $\beta$ and for small $\alpha$ is between $2\beta$ and $3\beta$
[cells/year] (see supplement for further discussion on this point).
If the expansion is to cover the site in the
interval $[L,U]$ then we need
\[\E(\beta)\gtrsim \frac{\max(C_1,C_2)}{2(L-U)}.\] Rate parameters $\alpha$ and $\beta$ are
otherwise positive so the maximum entropy priors are
\begin{equation}
\label{eq:alphabetaprior}
\alpha\sim\Exp(A/N(U-L)), \quad \beta\sim\Exp(B\max(C_1,C_2)/2(L-U)).
\end{equation}
Prior simulation showed $1\lesssim A \lesssim 20$ and $0.5 \lesssim B \lesssim 2$
generated a range of onset-fields with plausibly varied temporal ``roughness'' (increasing with $\alpha$)
and ``peak to valley depth'' (decreasing with $\beta$).

The onset-field process has a boundary effect. In realizations of the process, boundary cells are
typically occupied later than cells in the interior.
Since $$\int_{(-\infty,\psi_M]^C} p(\phi|\alpha,\beta,\psi) d\phi=1,$$ we can differentiate
with respect to $\alpha_c$ to get the moment identity
\begin{equation}\label{eq:ephi}
\E(\phi_c)=\psi_M-\E\left(\frac{1}{\alpha_c+\sum_{c'\in \N(c)}\beta_{c,c'}\mathbb{I}_{\phi_{c'}>\phi_c}}\right).
\end{equation}
Since corner and side cells have fewer neighbors than interior cells, the sum $\rho(\phi_{\N^+(c)})$
in the denominator \Eqn{eq:ephi} is distributed over a smaller total there. Differentiating again with respect to $\alpha_{c'}$ gives
\[\E\left(\left(\phi_c+\frac{1}{\rho(\phi_{\N^+(c)})}-\E\left(\phi_c+\frac{1}{\rho(\phi_{\N^+(c)})}\right)\right)
     \left(\phi_{c'}+\frac{1}{\rho(\phi_{\N^+(c')})}-\E\left(\phi_{c'}+\frac{1}{\rho(\phi_{\N^+(c')})}\right)\right)\right)=0,\]
so $\phi_c+\rho(\phi_{\N^+(c)})^{-1}$ and $\phi_{c'}+\rho(\phi_{\N^+(c')})^{-1}$ are uncorrelated,
and we have verified this numerically, as a check on our code.
The field of values $\phi_c+\rho(\phi_{\N^+(c)})^{-1}$ is homogeneous and isotropic to 2nd order,
regardless of boundary effects. Further field identities are given in
\Sec{sec:onsetalgsec} of the Supplementary material.
%

We might pad the
site with a large border outside the area where pits are concentrated, or, considering
\Eqn{eq:ephi}, raise $\alpha_c$ for $c$ the index of a boundary cell.
However, simulation of $p(\alpha,\beta,\phi|\psi)$ showed that these approaches
give a distribution of onset fields which does not represent
prior belief for this site. If there are just a few immigration events, and the
field is extended a great deal, then the immigration events tend to occur
in the extension, and the pit-area is settled by migration.
In this class of models, the interval between the onset event at time $\psi_M$
(people arrive on the beach) and occupation at cells covering pits (the settlement
reaches the excavated area) can be improbably large. By accepting
some spatial inhomogeneity in the prior onset-field at the boundary
we get a prior distribution better resembling prior belief.

Pit locations are a second source of spatial inhomogeneity.
The location of a pit with at least one dated specimen is informative for the onset field, without the
radiocarbon date itself,
because the deposition process must act for some finite time wherever there are dated specimens.
This constrains $\phi_{\c(x_h)}>\psi_0$ for $h=1,2,\dots,H$, but not elsewhere.
We go further and assert as part of the observational data
that the deposition process acts for some finite
time at each point over the windowed region of this site, so $\psi_0<\phi_c\le\psi_M$.
Conditional on this assertion, the locations of the dated pits are not informative.
As a side-effect, the marginal prior distribution of the span $\psi_M-\psi_0$ depends
on the priors for $\alpha$ and $\beta$. Short spans are excluded
at small $\beta$ as the field needs time to cover the site. As a second side effect,
the boundary effect mentioned in the previous paragraph is slightly reduced.

\begin{figure}[htb]
  \hspace*{-0.75in}\includegraphics[width=3.5in]{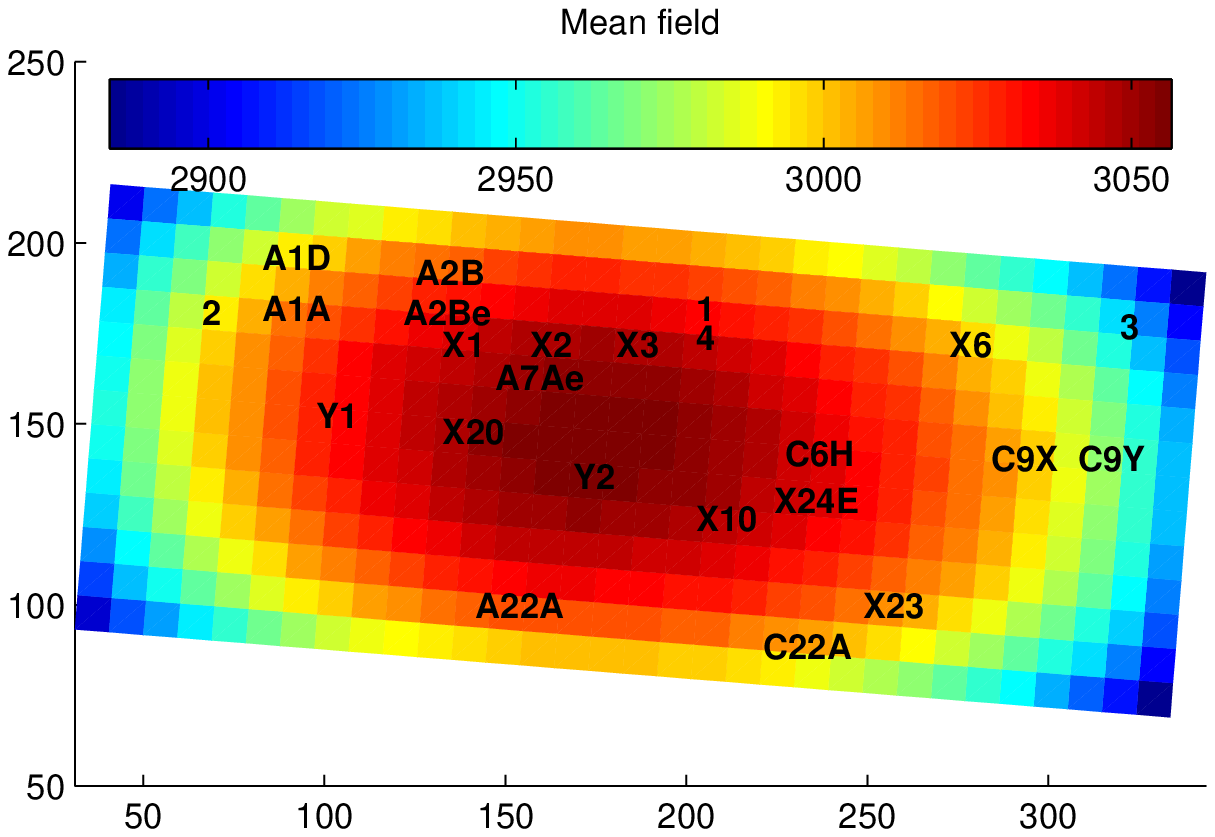}\includegraphics[width=3.5in]{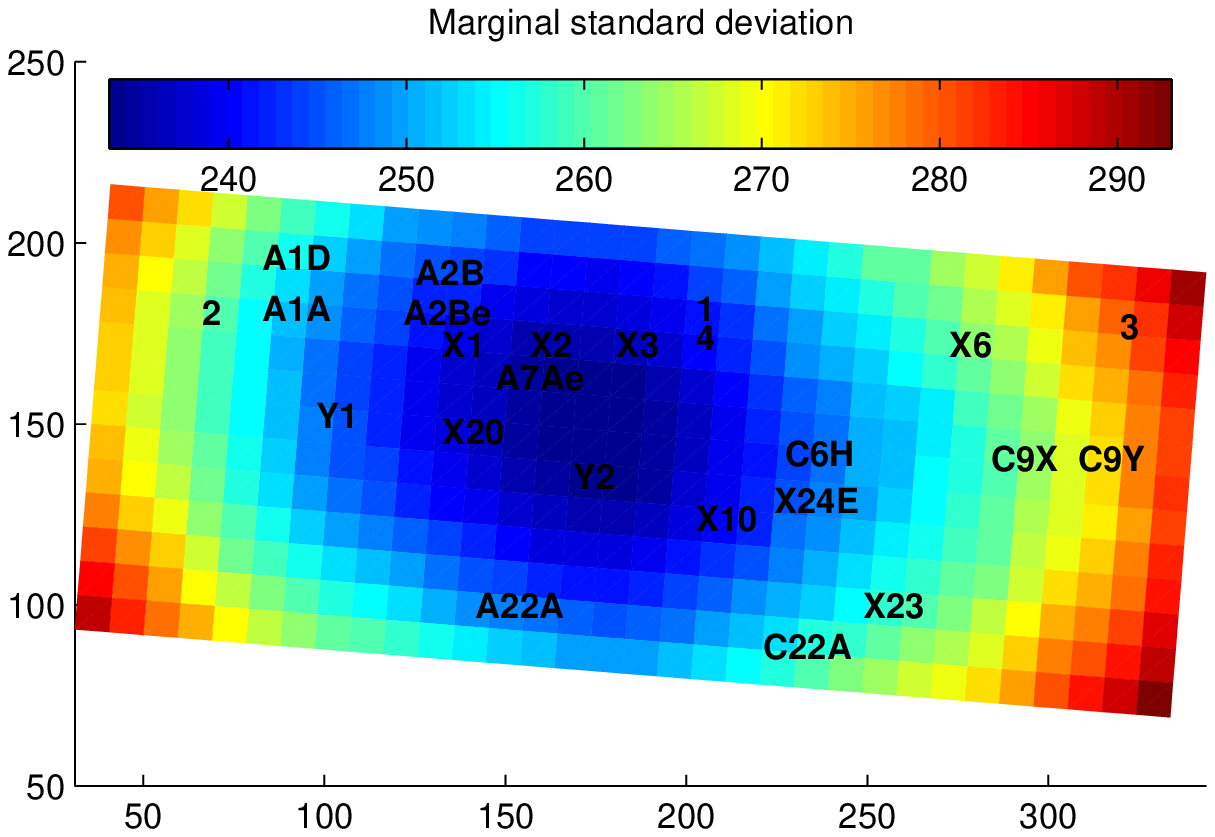}\\
  \vspace*{-0.25in}
  \caption{The prior mean onset field (left) and the prior standard deviation of the field (right),
    in the single-phase onset-field model of \protect\Sec{sec:onsetfieldbourewa}.
    The colors in the color bar give years BP.}\label{fig:priormeanstdNPYp0N11N60}
\end{figure}

A single sample from the prior $p(\theta,\psi,\phi,\alpha,\beta)$ given above,
for $M=1$ is shown in \Fig{fig:priorrealizationNPYp0N11N60}. The lattice is $32 \times 13$
and $A=10$ and $B=1$.
Cells are 2.375 meters on each side, and the field is piecewise constant by cell.
In \Fig{fig:priormeanstdNPYp0N11N60}
we show the prior mean and standard deviation
of the field estimated marginally for each cell. Further figures illustrating the onset-field prior
and the priors for $\psi$ and $\theta$ are shown in the Supplement. We see residual inhomogeneity in the prior of about 150
years from centre to corner. The marginal prior standard deviation of $\phi_c$ at cell $c$ is of order 250 years.
This includes random offset variation from $\psi_M$, the initialisation-time.
The standard deviation of the elapsed time $\psi_M-\phi_c$ is of the order of 200 years.

\section{Fitting the onset-field for a single phase}
\label{sec:firstfit}

The posterior density for deposition in a single phase ($M=1$), modeled as \Sec{sec:phasemodel},
with the onset-field of \Sec{sec:onsetfield} conditioned to satisfy $\phi\in [\psi_0,\psi_M]^C$
is
\[
p(\alpha,\beta,\phi,\psi,\theta|y,\m,M)\propto
\ell(\theta;y)p(\alpha,\beta,\phi,\psi,\theta|\m,M).
\]
For $p(\alpha,\beta,\phi,\psi,\theta|\m,M)$ see the definitions at \Eqn{eq:fullprior}.

We fit this model by Metropolis-Hastings MCMC simulation. The updates are those of \cite{nicholls01}.
New parameters $\alpha$ and $\beta$ are updated via a scaling proposal
$z\sim U(1/2,2)$, $\alpha'=z\alpha$, $\beta'=z\beta$
and jointly with the onset-field
in an update with proposal distribution $p(\phi|\psi)p(\alpha,\beta)$.
The algorithm is effective when the latter, which updates $\alpha$, the two components of
$\beta$ and the field $\phi$, has a reasonable success rate. We have a
second onset-field update with proposal distribution $p(\phi|\alpha,\beta,\psi)$,
which conditions on the immigration migration rate parameters.
The two $\phi$-updates generate new onset-field values at every cell, and are simulated using
the algorithm of \Sec{sec:onsetalgsec}. The
$\phi$ field has a large spatial correlation, with
structure on the same scale as the site itself, so single-cell MRF updates
proved to be very inefficient.
The condition $\phi\in [\psi_0,\psi_M]^C$ is implemented by rejecting
proposals at the acceptance stage of the MCMC update which do not satisfy the condition,
rather than by making proposals from the conditional distribution.

\begin{figure}[htb]
  \hspace*{-0.75in}\includegraphics[width=3.5in]{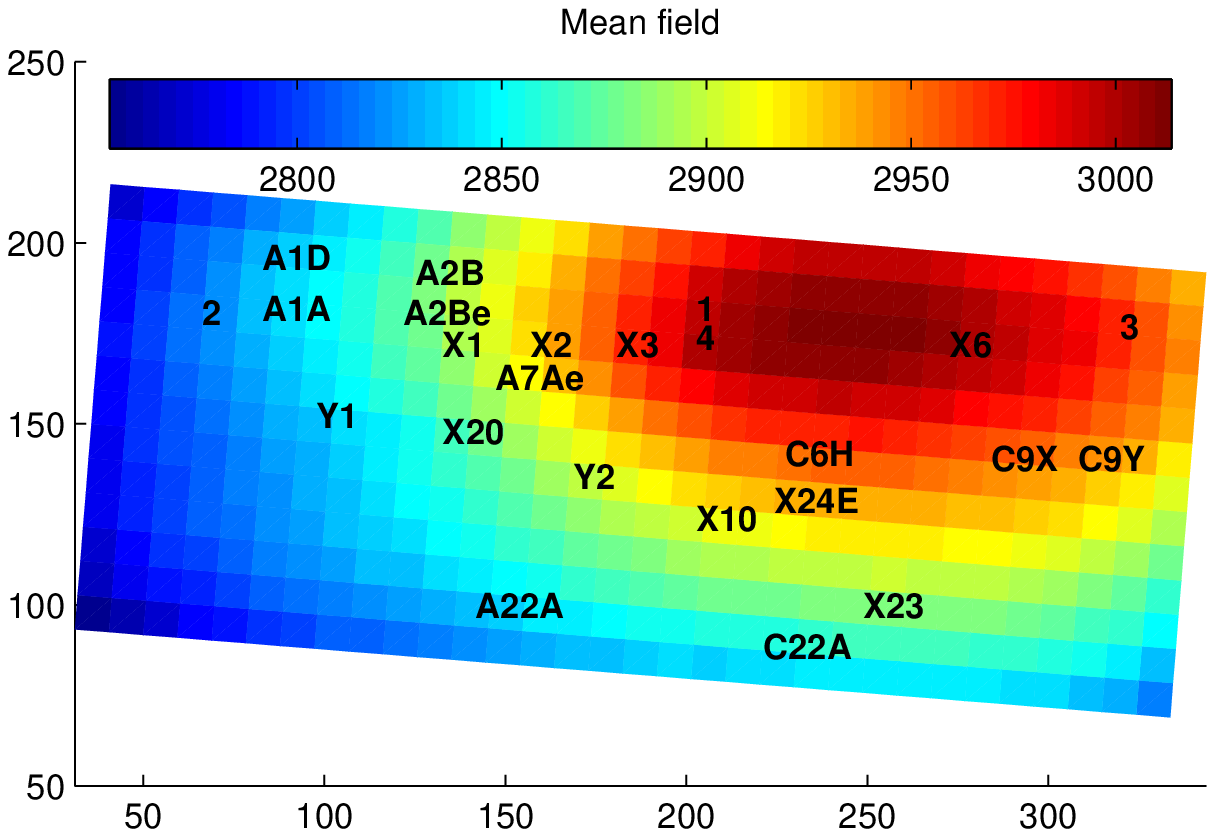}\includegraphics[width=3.5in]{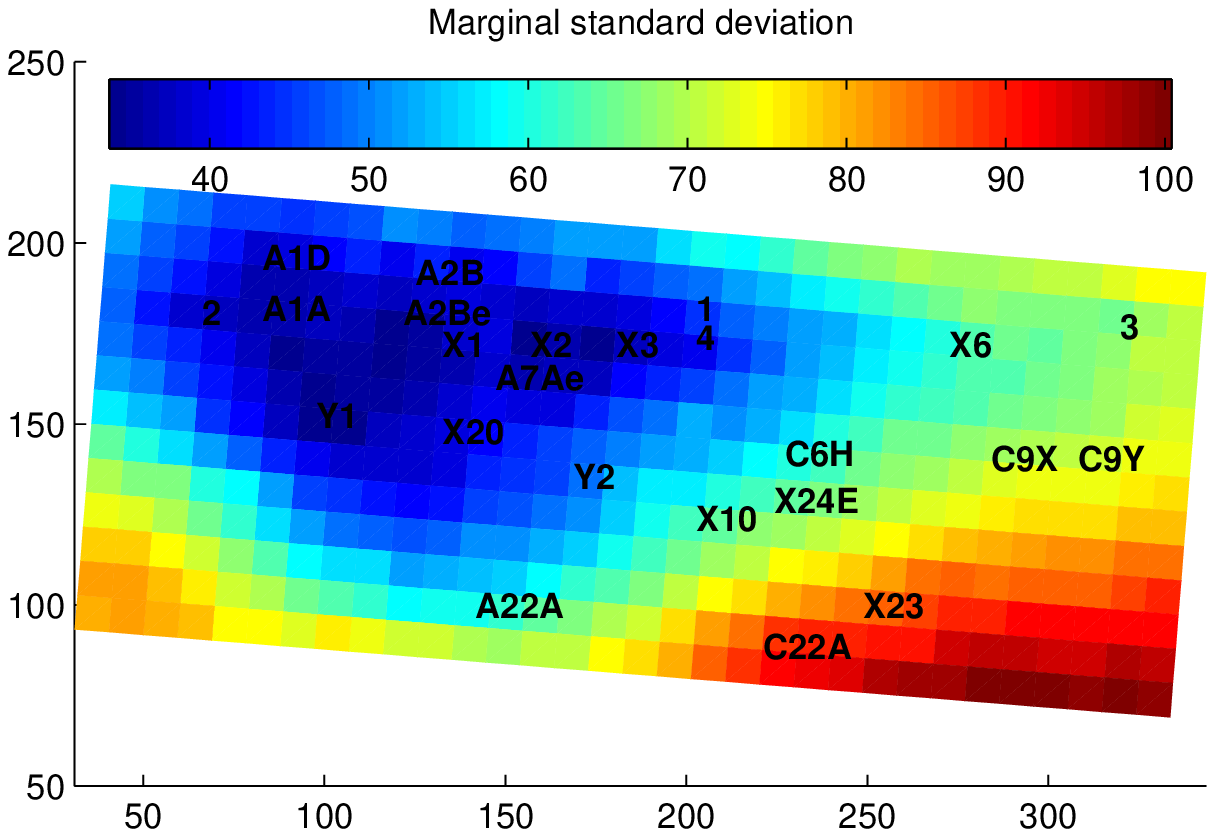}\\
  \vspace*{-0.25in}
  \caption{The posterior mean onset field (left) and the posterior standard deviation of the field (right),
    in the single-phase onset-field model of \protect\Sec{sec:firstfit}.}\label{fig:postmeanstdNPYp0N11N60}
\end{figure}

\begin{figure}[htb]
  \hspace*{-0.5in}
  \includegraphics[width=3.25in]{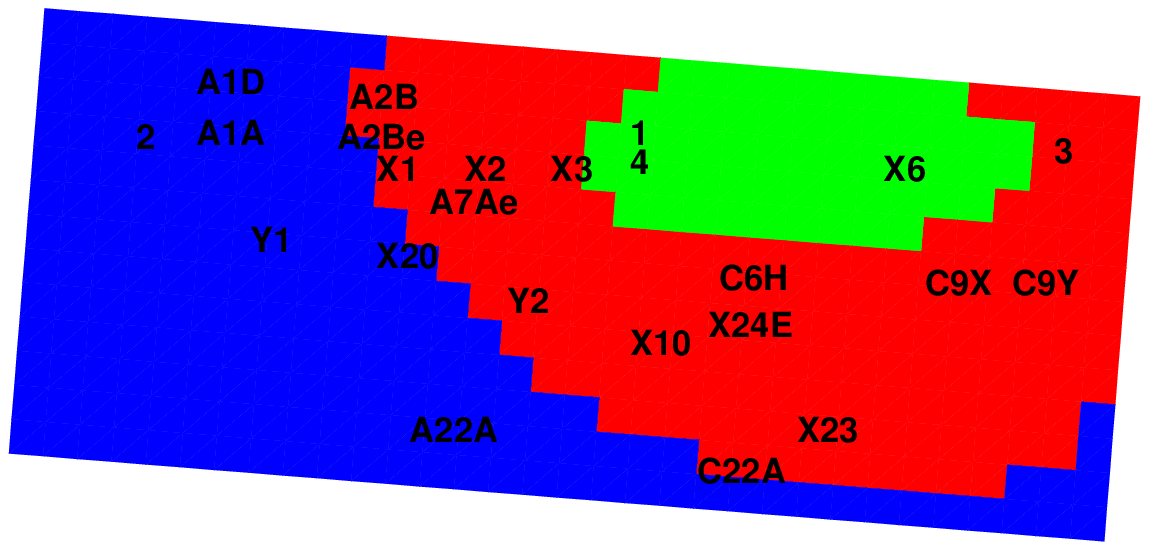}\quad\includegraphics[width=3.25in]{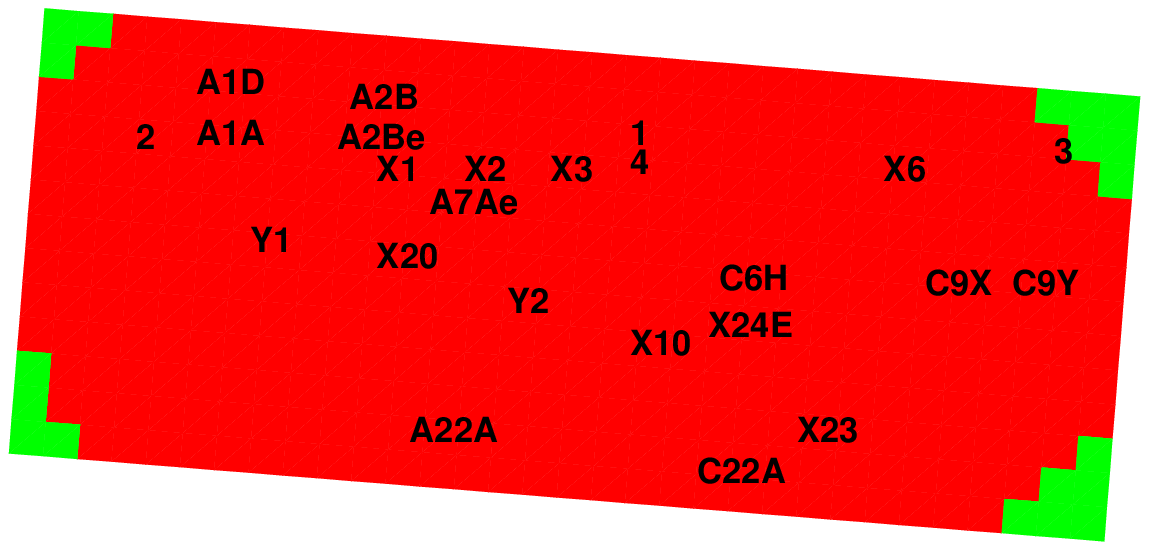}\\
  \caption{Support for spread: (left) the posterior probability that $\psi_M-\phi_c<150$
  (respectively, $\psi_M-\phi_c>150$)
  is greater than 0.8 for cells $c$ in the green (blue) region, in the single-phase onset-field model of \protect\Sec{sec:firstfit}.
  (right) the corresponding image for the prior distribution of the onset field.
  Results for the multi-phase onset-field model of \protect\Sec{sec:fullmodel} are similar.}
  \label{fig:fieldprobabilitypartition}
\end{figure}

An image of the posterior mean onset-field is shown in \Fig{fig:postmeanstdNPYp0N11N60}.
There
is a single peak. Compared to the prior field, \Fig{fig:priormeanstdNPYp0N11N60},
the range of onset times (the depth of the field) has gone up,
and the standard deviation, shown in the same figure, has dropped.
Typical values (across cells) of the standard deviation of the elapsed times $\psi_M-\phi_c$ drop from
values of order 160 years in the prior to values of order 60 years in the posterior.
\Fig{fig:pitages} shows the
posterior distribution for first settlement, $\max(\phi)=\psi_M$ (with $M=1$, leftmost red histogram)
at around 3100 BP. By contrast the single phase analysis puts this event
at around 2900 BP, since it is pulled down by pits with younger dates.
\Fig{fig:fieldprobabilitypartition}
shows the sets of cells
\begin{eqnarray}
    \label{eq:partition}
  \mbox{green} &=& \{c: \Pr(\psi_M-\phi_c>T^*|y)>p^*, c=1,2,\dots,C\}, \nonumber\\
  \mbox{blue} &=&  \{c: \Pr(\psi_M-\phi_c<T^*|y)>p^*, c=1,2,\dots,C\}, \\
  \mbox{red} &=& \{1,2,\dots,C\}\setminus (\mbox{green}\cup \mbox{blue}),\nonumber
\end{eqnarray}
with $T^*=150$[years] and $p^*=0.8$ and we estimate the probabilities using
MCMC simulation of $p(\alpha,\beta,\phi,\psi,\theta|y,\m,M)$.
Green cells are all settled within 150 years of first arrival,
and blue cells are all settled more than 150 years after
first arrival, with marginal cell probabilities all at least 0.8.
A visually interesting threshold $T^*$ was set by searching,
so there is a hazard akin to multiple testing. However, when we repeat this
exercise on prior simulations no single threshold $T^*$ makes both
'green' and 'blue' non-empty at $p^*=0.8$.

Independent pit-by-pit estimation of the onset and terminal events for a
phase model of the kind described in \Sec{sec:phasemodel} ($ie$, without onset field) is usually pointless
if there are no more than two dated specimen in each pit. The marginal posterior density for
the onset event has a tail which decays like $1/\psi_M$. As a consequence, results are sensitive
to the choice of the bound $U$. This is unsatisfactory, as $U$ is usually just a
very conservative upper bound on $\psi_M$, the first-onset. The blue histograms in \Fig{fig:pitages}
show these pit-by-pit analyses.
The red histograms in \Fig{fig:pitages}
show the posterior distributions of $\phi_{\c(x(h))}$, the age of the onset-event
at pit $h=1,2,\dots,H$. The onset field spatially smooths the onset times in the pit-by-pit analysis,
concentrating the onset time distributions within the support of the pit-by-pit distributions.

Where the red onset-field and blue single-pit histograms in \Fig{fig:pitages}
actually conflict, there is possible
evidence for model mispecification in the onset field. This occurs at pits `4' and `X20'.
Because these pits have just two dates each, the evidence for mispecification
is not compelling. In pit `X20' we have two consistent, relatively late, dates
(see \Fig{fig:bourewadata}) while the onset field interpolates an earlier onset.
However, the red and blue histograms at `X20' do overlap a little, reflecting the
fact that we may by chance have chosen a couple of late specimens from a population
of specimens in accord with the onset field estimate.

We are assuming that there was no abandonment of areas once settled, up to $\psi_0$,
the end of deposition associated with the culture of interest.
The lower plot in \Fig{fig:pitages} shows the all-in-one phase (green)
the pit-by-pit (blue) and onset-field (red) estimates for $\psi_0$.
The single-pit analysis in pit `4' favors an earlier abandonment.
Looking at \Fig{fig:bourewadata} we see two early dates and no later dates,
so this might be abandonment. On the other hand pit `4' is adjacent pit `1',
where there is no evidence for abandonment.

\begin{figure}[htb]
  \[\hspace*{-0.25in}\includegraphics[width=5in]{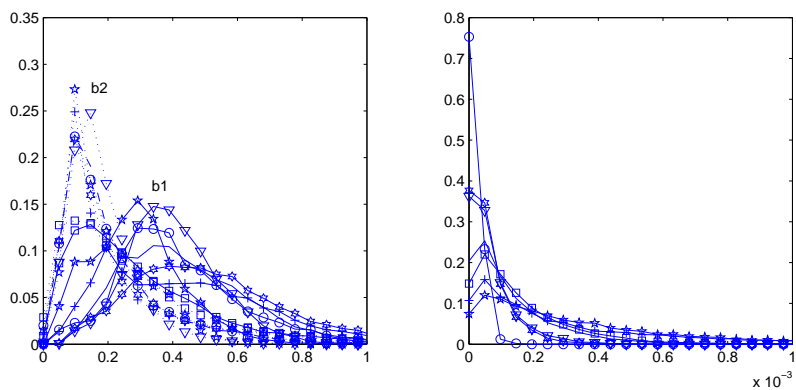}\]
  \vspace*{-0.2in}
  \caption{$x$-axis 25cm units/year ($ie$, rate multiplied by cell side length),
        $y$-axis frequency. (Left) the posterior distributions of migration rate
            $\beta_1$ (solid, `b1') and $\beta_2$ (dashed, `b2')
            in six independent single-phase onset-field models with the $A$ and $B$
            hyperprior parameters of \protect\Eqn{eq:alphabetaprior}
            given by $(A,B)=(10,1)$ (no symbol), $(A,B)=(20,2)$ (+),
            $(A,B)=(5,0.5)$ (triangle), $(A,B)=(20,0.5)$ (5-star), $(A,B)=(5,2)$ (6-star) and $(A,B)=(10,1)$
            on a $64\times 25$ lattice (circle). Prior simulation at $(A,B)=(10,1)$ has square-symbols.
            (right) corresponding posterior distributions for $\alpha$.}
            \label{fig:abplot}
\end{figure}

We have varied the priors
for the immigration $\alpha$ and migration-speed $\beta$ parameters which control the
shape of the onset field, and checked that results $\alpha$ and $\beta$ are robust to reasonable variation.
The posterior distribution of the migration rates $\beta_1$ and $\beta_2$
and immigration rate $\alpha$ for model fitting with different priors,
scaling the prior mean rates by 2, and by 1/2, are shown in \Fig{fig:abplot}.
The prior simulations (left in \Fig{fig:abplot}, colored green,
solid and dashed respectively) for $\beta_1$ and $\beta_2$ coincide.
The distributions
of the two rates differ in the posterior, and are not shifted by the changing priors
with doubled and halved rates.
Expansion of the settlement moved more rapidly along than up the beach.
The immigration rate is more sensitive to the prior within the range considered.

In order to measure support for spread from a single centre,
we count the number of arrival-events,
$$V(\phi)=\sum_{c=1}^C\mathbb{I}_{\phi_c>\phi_{c'}\forall c'\in \N(c)}$$
in a field. For example,
the prior sample in \Fig{fig:priorrealizationNPYp0N11N60} has $V=2$ arrivals.
The posterior to prior odds for $V=1$ in a run with $A=10$ and $B=1$
are greater than one and the odds decline steadily with increasing $V$, so the data
push the distribution of $V$ towards fewer arrival events.
Our analysis of the model with hyper-prior parameters $(A,B)=(10,1)$ on a $64\times 25$-cell lattice
gave essentially identical results. The Bayes factor for $V=1$ against $V=2$ is only slightly
greater than one. Very small peaks mask the big picture of \Fig{fig:fieldprobabilitypartition}.
Supporting figures are presented in the Supplementary material.

We have checked that the effects we see are not imposed by the prior distributions
we are using. We made a single run with the
lattice size doubled to $64\times 25$ on the same region. Results for the default parameter values
$A=10$, $B=1$ were very similar to those above (see Supplement).
We simulated synthetic data under the single-phase
model of \Sec{sec:phasemodel} and fit it with the single-phase onset field model.
We do not detect a settlement process where there is none. There is no threshold
that splits the cells at $p^*=0.8$, that is, one of the green or blue sets is empty
for each threshold $T^*$. Also,
the posterior to prior odds are flat with increasing $V$. Supporting figures are presented in the Supplementary material.

\section{Recovering phase structure without spatial structure}
\label{sec:phasestructure}

We now extend our analysis to the case where the number of phases,
and the assignment of specimen to phases, is not known. In this section
we make a purely temporal change-point analysis for the unknown piecewise-constant
dated-specimen deposition rate.



\subsection{Prior model for assignments of specimens to phases}
\label{sec:phasestructuremodel}
The number of phases $M$, and the map $\m$ from data index to phase,
are now random variables.
For $m=1,2,\dots,M$,
let $K_m(\m)=\card\{i: \m(i)=m, i=1,2\dots,K\}$ give the number of dates in the $m$'th phase,
and
\[
p_m(\lambda_\theta,\psi)=\frac{\lambda_{\theta,m}(\psi_{m}-\psi_{m-1})}
{\sum_{m'=1}^M\lambda_{\theta,m'}(\psi_{m'}-\psi_{m'-1})}.
\]
give the probability for specimen $i$ to be assigned to phase $\m(i)=m$.
The prior probability for
assignment $\m$ is a function of ratios of the unknown
dated-specimen deposition rates $\lambda_1,\lambda_2,\dots,\lambda_M$,
\begin{equation}\label{eq:mgivenlambdapsi}
p(\m|\lambda_\theta,\psi)=\prod_{m=1}^M p_m(\lambda_\theta,\psi)^{K_m(\m)}
\end{equation}
The right hand side of \Eqn{eq:mgivenlambdapsi} has a multinomial form, without the
$K!/\prod_m K_m!$ factor, because the left hand side is the probability for
an assignment $\m$ of distinct specimen to phases and there are $\prod_m K_m!/K!$ such assignments.
The posterior distribution is now
\begin{equation}\label{eq:standardVDpost}
p(M,\psi,\lambda_\theta,\m,\theta|y)\propto
\ell(\theta;y)p(M,\psi,\lambda_\theta,\m,\theta),
\end{equation}
with
\[
p(M,\psi,\lambda_\theta,\m,\theta)=p(M)p(\psi|M)p(\lambda_\theta|M)p(\m|\lambda_\theta,\psi)p(\theta|\psi,\m).
\]
If we condition on $M$ and $\m$ in \Eqn{eq:standardVDpost} 
we get \Eqn{eq:standardpost}.
Of these factors, $p(\theta|\psi,\m)$ and $p(\psi|M)$ are given by \Eqns{eq:psigivenM}{eq:thetagivenpsim}
and $p(\m|\lambda_\theta,\psi)$ in \Eqn{eq:mgivenlambdapsi}. The prior distribution
of $M$ (in fact $M-1$, the number of phase boundaries between onset and ending)
is Poisson. If we took our prior straight from the deposition-process $\Lambda_{\Psi}$ of \Sec{sec:phasemodel} we would
have $M-1\sim\mbox{Poisson}(\lambda_\psi(\psi_M-\psi_0))$. In fact
we take $M-1$ Poisson distributed, but fix the mean at $\log(2)$ so
that the prior probability that $M=1$ is equal to one half.

Our prior for the dated-specimen deposition rates $\Lambda_{\theta}|M$ is $\lambda_{\theta,m}\sim\Exp(1)$ for
$m=1,2,\dots,M$. The common scale of the rates is irrelevant as the rate-prior
is important only insofar as it decides the prior for the phase allocation probabilities
$(p_1,p_2,\dots,p_M)|M$. In our setup, for given $\psi$, dated specimens are {\it a priori} more likely to belong to long-lasting phases.
When $\psi_{m}-\psi_{m-1}$ are equal for $m=1,2,\dots,M$, we get
$(p_1,p_2,\dots,p_M)|\psi\sim\mbox{Dirichelet}(1,1,\dots,1)$.
Simulation of the prior shows that, allowing for variation in both $\psi$ and $\lambda_\theta$,
$(p_1,p_2,\dots,p_M)|M \sim \mbox{Dirichelet}(1/2,1/2,\dots,1/2)$
is a good approximation, for at least $M\lesssim 10$.
In the supplement we show that, if we accept this Dirichelet approximation,
then the array of numbers of dates in phases, $(K_1,K_2,\dots,K_M)|M$, has a multivariate Polya distribution,
\[
p(K_1,K_2,\dots,K_M|M)\simeq \frac{K!}{\prod_{m=1}^M K_m!}\frac{\Gamma(M/2)}{\Gamma(K+M/2)}
               \prod_{m=1}^M \frac{\Gamma(K_m+1/2)}{\sqrt{\pi}}.
\]
\cite{mosimann62} gives the moments. The mean is $\E(K_m)=K/M$, as we expect.
The covariance matrix is a scalar multiple of the covariance matrix of a
$\mbox{Multinomial}(1/M,\dots,1/M)$ array and the two distributions have equal correlation
matrices. We conclude that the simple choice $\lambda_{\theta,m}\sim \Exp(1)$ seems to lead
to no unacceptable marginal prior distributions, and other aspects of the prior framework
are motivated by the deposition-model framework.

\subsection{Estimated phase structure for the Bourewa and Stud Creek data sets}
\label{sec:fitphasestructure}

We fit the random phase structure model of \Sec{sec:phasestructuremodel} using MCMC.
We extend the updates of \cite{nicholls01} to include moves to add and delete phase boundaries,
and to move specimen between phases (by varying a specimen age or a phase boundary age).
When we add a phase boundary into an existing phase, the state dimension increases by
two, as we need a deposition rate for the new phase. We have the Metropolis Hastings Green
setup for MCMC in something very like the original variable-dimension application of \cite{green95}.

\subsubsection{Bourewa}
\label{sec:fitphasestructurebourewa}
When we fit this model to the Bourewa data,
we are ignoring spatial structure. The posterior mode has 2 phases.  When we plot
the locations
of specimens assigned in the mode to phases one to three on separate graphs,
we get a picture which resembles
\Fig{fig:agesandscatter}. The plot is omitted, for brevity, and because it is misleading,
as we now explain.

The analysis seems to show
expansion from a centre, with the distribution in specimen locations spreading out
as we move from phase to phase forwards in time. Do we need an onset-field analysis
if the phase assignment model shows such a clear effect?
First, and most important, central pits have more dates,
so they may display more early dates. Secondly,
if the settlement did expand, and the specimen are selected for dating uniformly
in space and time, then we get more dates when the settlement is larger,
and we get a good fit to the rate with a staircase of physically spurious phases.

\subsubsection{Stud Creek}
\label{sec:fitphasestructurestud}
The phase-assignment analysis is useful for measuring support for
alternative temporal phase-structure models, as we now illustrate.

\begin{figure}[htb]
  \includegraphics[width=6in]{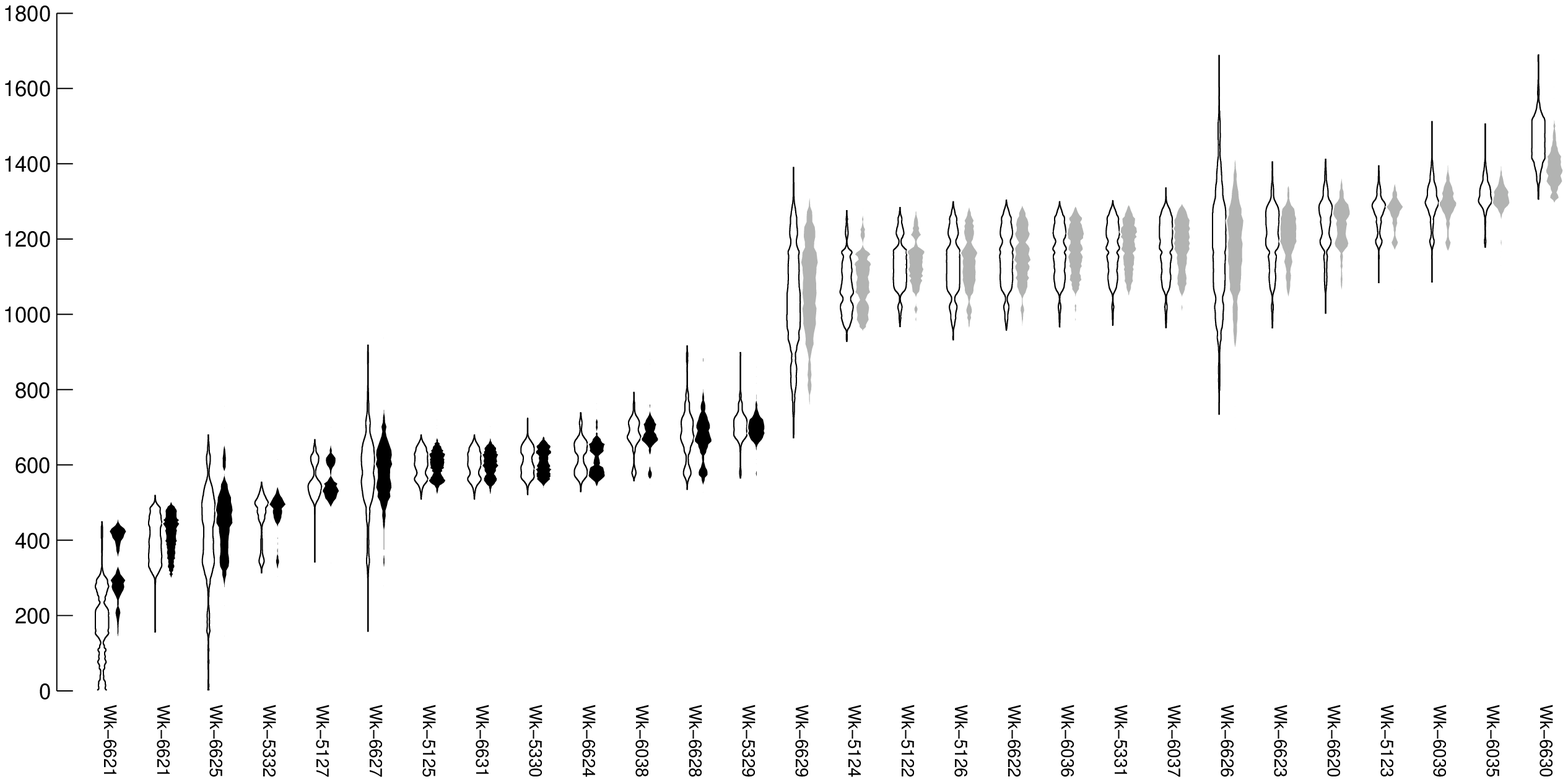}\\
  \caption{(black outline) likelihoods for the Stud Creek data of \protect\cite{holdaway02};
  (black, gray filled) marginal posterior
  distributions of ages shaded by phase assignment (black) phase 1, and (gray) phase 3
  and conditioned on that assignment.
  (y-axis) years BP, (x-axis) specimen label per
  \protect\cite{holdaway02}.}\label{fig:stud}
\end{figure}

Stud creek is a small stream system in far western New South Wales, Australia.
The area was excavated between 1996 and 1998, and the remains of 72
hearths were recorded, of which 28 yielded one dated specimen each. These data are reported in
\cite{holdaway02} and \cite{holdaway08},
who argue that, in our terms, the dated hearths are a random sample from the
population of previously buried hearths in the study area.
Although there has been some localized erosion,
hearths are scattered over a wide area, with hearths of markedly different ages located
adjacent to one another. Their likelihood is graphed in \Fig{fig:stud}.
The archaeologists conject that the site was in use in two separate phases, with a hiatus
associated with climate change.
This question is answered in \cite{holdaway02}. We use these data to illustrate
our methods for recovering archaeological phase structure. Exploratory analysis by the
authors of \cite{holdaway02} showed no spatial structure in the specimen ages.

In order to estimate the phase structure of the Stud creek data we fit
the phase-structure model of this section to the data (with limits
$L=0$ and $U=2000$ years BP). When we fit a phase structured model
to determine phase, we are making a model comparison over phase models in a discrete set.
The prior distribution over phase models allows any number $M$ of phases, and
an arbitrary assignment $\m$ of specimen to phases. In the past, model comparison
of this kind has been
made for a very small number of alternative models, typically, two.
For example, \cite{jones02} estimate Bayes factors in order to weigh support
for two models of a small radiocarbon date data set with ($M=6, \m=(1,2,3,4,5,5,6)$) and without ($M=1, \m=(1,1,1,1,1,1,1)$)
stratigraphic phases, from $K=7$ dated specimen.
For the Stud creek data, \cite{holdaway02} compare
model $A=\{M=3, \m=(1\times 13,3\times 15)\}$ with model
$B=\{M=1, \m=(1\times 28)\}$ for the $K=28$ stud creek dates. Model A has an empty phase
(phase $2$) for hearth construction explaining the obvious step in \Fig{fig:stud}.
We will make two model comparisons, between models $A$ and $B$,
and between models $A'=\{M=3\}$ and $B'=\{M=1\}$. The $A'\ v.\ B'$ comparison
weighs the evidence for three phases, as opposed to one, without specifying
which specimens are in which phase. 
The models in the $A\ v.\ B$ comparison specify specimens' phase assignments.

We begin with the $A'\ v.\ B'$ comparison.
The Bayes factor
is the ratio of the posterior to prior odds ratios,
\[
{\mathcal B}(A',B')=\frac{\Pr(A'|y)}{\Pr(B'|y)}\frac{\Pr(B')}{\Pr(A')}.
\]
We estimate $\widehat{\Pr(A'|y)}\simeq 0.20$ and $\widehat{\Pr(B'|y)}\simeq 0.41$
as the proportion of states $X_t$ with $M=3$ and $M=1$ phases respectively,
in MCMC simulations with $X^{(t)}=(M^{(t)},\psi^{(t)},\lambda_\theta^{(t)},\m^{(t)},\theta^{(t)})$
for $t=1,2,\dots,T$ and convergence $X^{(t)}\stackrel{D}{\rightarrow}p(M,\psi,\lambda_\theta,\m,\theta|y)$
to the posterior. The quantities $\Pr(A')\simeq 0.12$ and $\Pr(B')=0.5$
are $\mbox{Poisson}(M-1;\log(2))$. We find ${\mathcal B}(A',B')\simeq 2.03$
so there is very mild support for 3 phases over one.
The reason for this weak positive support, is that model $A'$ is a large class of models,
including many very improbable assignments $\m$ of specimens to its three phases.
The mean likelihood in the $A'$-prior is lowered by these models.

Since there is just one way to assign $K$ specimen to $M=1$ phases,
$B=B'$ and we have $\widehat{\Pr(B|y)}\simeq 0.41$ and $\Pr(B)=0.5$
as before. The posterior
probability for $A$ was large enough to estimate simply, at $\widehat{\Pr(A|y)}\simeq 0.08$.
The prior for $A$ is
$\Pr(M,\m)=\Pr(\m|M)\Pr(M)$ with $\Pr(M=3)\simeq 0.12$ and $P(\m|M)$ calculated in
the supplement as $\Pr(\m|M)\simeq 10^{-10}$.
The Bayes factor is then ${\mathcal B}(A,B)\simeq 8\times 10^9$ so the support for
$A$ over $B$ is overwhelming. The red and blue histograms in \Fig{fig:stud}
show the posterior distribution for the hearth ages, conditioned on model A,
whilst the colors show the maximum a posteriori phase assignment.

Inference for $M,\m$ is useful for exploratory analysis.
When we fit the Stud-creek data, in the $(A',B')$-analysis, the hiatus model $A$ is thrown up as worth
closer inspection in a search over many models.
We looked at other priors on $M$ and $\m$: one of these
is discussed further in the supplement.

\section{Fitting the onset-field for multiple phases}
\label{sec:fullmodel}

If a settlement forms, spreads, and from time to time undergoes dramatic changes
in character, so that the deposition at all settled locations moves into a new phase,
we should fit a model with local deposition switched on by the spreading onset-field,
cut across with an unknown number $M$ of site-wide phase boundaries.
This is the onset-field model described in \Sec{sec:onsetfieldbourewa} augmented with the prior
weighting for $M$, the relative deposition rates
$\lambda_\theta=(\lambda_{\theta,1},\lambda_{\theta,2},\dots,\lambda_{\theta,M})$
and the specimen to phase map $\m$ described in \Sec{sec:phasestructure}.
The posterior density is
\[
p(\alpha,\beta,\phi,M,\psi,\lambda_\theta,\m,\theta|y)\propto
\ell(\theta;y)p(\alpha,\beta,\phi,M,\psi,\lambda_\theta,\m,\theta),
\]
with
\[
p(\alpha,\beta,\phi,M,\psi,\lambda_\theta,\m,\theta)=
p(\alpha,\beta)p(M)p(\psi|M)p(\lambda_\theta|M)p(\phi|\psi,\alpha,\beta)
p(\m|\lambda_\theta,\psi,\phi)p(\theta|\psi,\m,\phi).
\]
In order to clarify the state, we explain how to simulate the prior.
We simulate $M\sim\mbox{Poisson}(\log(2))$ and $\psi$ according to \Eqn{eq:psigivenM}.
The onset field is simulated using the algorithm in \Sec{sec:onsetalgsec},
with $\alpha$ and the two components of $\beta$ independent and distributed as
\Eqn{eq:alphabetaprior}. These steps are repeated until $\max(\phi)\ge \psi_0$.
We have now to assign the specimen to phase map $\m$ and specimen ages
$\theta_1,\theta_2,\dots,\theta_K$.
The probability, $p_{m,i}$ say, for specimen $i$ to be assigned to phase $m$ depends now
on the local onset field, since the deposition rate at a pit before onset is zero.
Deposition in phase $m$ does not occur at pit $h$ at ages before $\phi_{c(x_h)}$.
Let
\[
\Delta_{m,h}=\max(0,(\min(\phi_{\c(x_{h})},\psi_{m})-\psi_{m-1}))
\]
give the span of deposition in phase $m$ at pit $h$. The conditional single-specimen
phase-assignment probability is
\[
p_{m,i}=\frac{\lambda_{\theta,m}\Delta_{m,\h(i)}}
{\sum_{m'=1}^M\lambda_{\theta, m'} \Delta_{m',\h(i)}}
\]
independently for each $i=1,2,\dots,K$ and the conditional probability for any particular phase
assignment $\m$ is
\[
p(\m|\lambda_\theta,\psi)=\prod_{i=1}^K p_{\m(i),i}(\lambda_\theta,\psi,\phi).
\]
Finally, the unknown true specimen ages $\theta_i,\ i=1,2,\dots,K$
are uniformly distributed in $\psi_{\m(i)-1}<\theta_i<\min(\psi_m,\phi_{\c(i)})$
as before.
The MCMC simulation of the posterior $p(\alpha,\beta,\phi,M,\psi,\lambda_\theta,\m,\theta|y)$ combines the
MCMC updates of Sections~\ref{sec:firstfit}~and~\ref{sec:fitphasestructure}. Each
MCMC simulation involving onset-fields takes at least 24 hours to run.

The posterior onset-field distribution we obtain fitting this second random-phase model
is very similar to the distribution we had in the single phase onset-field model.
The figures corresponding to
Figures~\ref{fig:priormeanstdNPYp0N11N60},~\ref{fig:postmeanstdNPYp0N11N60}~and~Figure~\ref{fig:fieldprobabilitypartition}
appear in the supplementary material.

The modal number of phases is one. However, the prior weights in favor
of a single phase. We can
estimate marginal likelihoods, $$e_m=\Pr(M=m|y)/\Pr(M=m)$$ for $M=m$ phases,
averaged over all other parameters. The ratio $e_{m_1}/e_{m_2}$ is the Bayes factor for
the model comparison of $M=m_1$ against $M=m_2$ phases. We find $\hat e_{1}\simeq 0.8$,
$\hat e_{2}\simeq 1$ and $\hat e_{3}\simeq 1.3$, so
the single phase model of \Sec{sec:firstfit} is not contradicted.

\begin{figure}[htb]
  \hspace*{-1.2in}\includegraphics[width=8in]{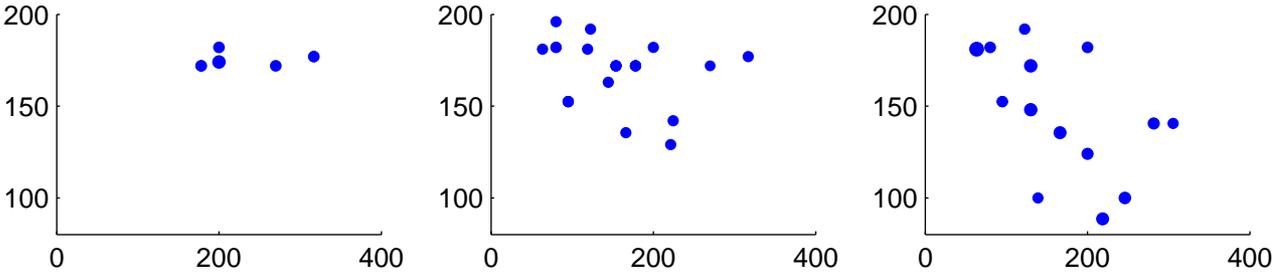}\\
  \vspace*{-0.1in}
  \caption{The scatter of specimen by phase, in the posterior mode assignment $\m$
  conditional on three phases, in a random-phase model with onset-field. (left) Phase 3
  (middle) Phase 2 (right) Phase 1. Axes are
  specimen excavation coordinates. Time
  increases left to right from phase 3 to phase 1. Point-size is proportional
  to posterior probability for the given phase assignment.}\label{fig:agesandscatter}
\end{figure}

In order to visualise the settlement directly on the data, we look at the distribution
of dates when there are three phases (supported by the largest reliably estimated marginal
likelihood). The scatter of the
specimens in each phase is shown in \Fig{fig:agesandscatter}.
The scatter plots shows a clear spreading trend as we move from phase to phase.

\section{Conclusions}

We have fitted four models to the data: a simple single-phase model without spatial structure
(\Sec{sec:phasemodel}), a single-phase model with an onset field (\Sec{sec:firstfit}),
a random-phase model with no spatial structure (\ref{sec:fitphasestructure}),
and a random phase model with onset-field (\Sec{sec:fullmodel}).
The single-phase onset-field (second model) models the unknown specimen dates
at each pit as uniformly distributed in time between the local onset and the end-of-deposition event.
The single-phase model without onset-field (first model) applies too much shrinkage to the earliest dates,
as the total deposition rate in a spreading settlement is not initially large.
This is visible in \Fig{fig:pitages} where the single-phase estimate for $\psi_M$
is shifted by 200 years towards the present. The single-phase onset-field model
estimate 3000-3200 BP will be more reliable. The onset-field priors do not
overwhelm the data.

In the random-phase model (third model),
phases are the pieces of the piece-wise constant specimen-age intensity.
Like the first model, this model treats the site as a single spatially unstructured
pit, ignoring variations in the intensity which might be due to onset-time varying from place to place.
It will detect the low initial deposition rate and not over-shrink. However,
it is unlikely that the phases recovered by the random phase model correspond to
physically distinct phases, in the archaeological sense. The evidence for phases
is variation in the intensity of the dated-specimen ages. Such variation might
be caused by any one of many time-varying selection processes which thin dateable material
down to dated specimen. However, the random phase model does provide, at
least heuristically, for rate variation, in the spirit of \cite{blaauw05},
who model the age-depth relation with a variable number of accumulation sections,
and \cite{karlsberg06}, who parameterises rate variation within a fixed number of phases.
The Stud Creek hiatus example has the unusual property that confounding processes may be absent.

The random-phase onset-field model (fourth model) adapts to heterogeneity in spatial and temporal
deposition rates. It is rather complex. However, it functions as a model-mispecification
check on the constant-rate hypothesis of the simpler single-phase onset-field model,
which it supports, as it allows a single phase.

Our analysis (\Fig{fig:pitages}) shows that the human-associated deposition did not commence at each pit location
at the same time. The posterior mean onset field \Fig{fig:postmeanstdNPYp0N11N60} shows the
pattern of expansion from a central location in the onset of deposition at Bourewa.
Ancient human settlement generating dated specimen begins in the area between pits
`1',`4' and `X6' and takes some 150-200 years to spread some 40-50 meters along the beach to pits `2', `A1A'
and `A1D'. 

\section*{Acknowledgements}
GKN acknowledges discussions with Tom Higham
of the Oxford Radiocarbon Dating Laboratory.
Most radiocarbon dates for this study were funded by the University of the South Pacific to PDN
who also thanks the Taukei Gusuituva and the people of the Bourewa area for their assistance
and hospitality.

\bibliographystyle{Chicago}
\bibliography{bourewa}

\begin{thebibliography}{}

\bibitem[\protect\citeauthoryear{Ammerman and Cavalli-Sforza}{Ammerman and
  Cavalli-Sforza}{1971}]{ammerman71}
Ammerman, A. and L.~Cavalli-Sforza (1971).
\newblock Measuring the rate of spread of early farming in {E}urope.
\newblock {\em Man\/}~{\em 6}, 674--688.

\bibitem[\protect\citeauthoryear{Bayliss, Ramsey, van~der Plicht, and
  Whittle}{Bayliss et~al.}{2007}]{bayliss07}
Bayliss, A., C.~B. Ramsey, J.~van~der Plicht, and A.~Whittle (2007).
\newblock {B}radshaw and {B}ayes: Towards a timetable for the {N}eolithic.
\newblock {\em Cambridge Archaeological Journal\/}~{\em 17\/}(1 (suppl.)),
  1--28.

\bibitem[\protect\citeauthoryear{Blaauw and Christen}{Blaauw and
  Christen}{2005}]{blaauw05}
Blaauw, M. and J.~Christen (2005).
\newblock Radiocarbon peat chronologies and environmental change.
\newblock {\em Appl. Statist.\/}~{\em 54}, 805--816.

\bibitem[\protect\citeauthoryear{Blackwell and Buck}{Blackwell and
  Buck}{2003}]{blackwell03}
Blackwell, P. and C.~Buck (2003).
\newblock The {L}ate {G}lacial human reoccupation of north-western {E}urope:
  New approaches to space-time modelling.
\newblock {\em Antiquity\/}~{\em 77}, 232--240.

\bibitem[\protect\citeauthoryear{Buck, Litton, and Smith}{Buck
  et~al.}{1992}]{buck92}
Buck, C., C.~Litton, and A.~Smith (1992).
\newblock Calibration of radiocarbon results pertaining to related
  archaeological events.
\newblock {\em Journal of {A}rchaeological {S}cience\/}~{\em 19}, 497--512.

\bibitem[\protect\citeauthoryear{Davison, Dolukhanov, Sarson, Shukurov, and
  Zaitseva}{Davison et~al.}{2009}]{davison09}
Davison, K., P.~Dolukhanov, G.~Sarson, A.~Shukurov, and G.~Zaitseva (2009).
\newblock Multiple sources of the {E}uropean {N}eolithic: Mathematical
  modelling constrained by radiocarbon dates.
\newblock {\em Quaternary International\/}~{\em 203}, 10--18.

\bibitem[\protect\citeauthoryear{Diggle, Rowlingson, and Su}{Diggle
  et~al.}{2005}]{diggle05}
Diggle, P., B.~Rowlingson, and T.~Su (2005).
\newblock Point process methodology for on-line spatio-temporal disease
  surveillance.
\newblock {\em Environmetrics\/}~{\em 16}, 423--434.

\bibitem[\protect\citeauthoryear{Durrett and Levin}{Durrett and
  Levin}{1994}]{durrett94}
Durrett, R. and S.~Levin (1994).
\newblock Stochastic spatial models: A users guide to ecological applications.
\newblock {\em Philosophical Transactions: Biological Sciences\/}~{\em 343},
  329--350.

\bibitem[\protect\citeauthoryear{Durrett and Liggett}{Durrett and
  Liggett}{1981}]{durrett81}
Durrett, R. and T.~M. Liggett (1981).
\newblock The shape of the limit set in {R}ichardson's growth model.
\newblock {\em The Annals of Probability\/}~{\em 9}, 186--193.

\bibitem[\protect\citeauthoryear{Graham, Lundelius, Graham, Schroeder, III,
  Andersona, Barnosky, Burns, Churcher, Grayson, Guthrie, Harington, Jefferson,
  Martin, McDonald, Morlan, Jr., Webb, Werdelin, and Wilson}{Graham
  et~al.}{1996}]{graham96}
Graham, R., E.~Lundelius, M.~Graham, E.~Schroeder, R.~T. III, E.~Andersona,
  A.~Barnosky, J.~Burns, C.~Churcher, D.~Grayson, R.~Guthrie, C.~Harington,
  G.~Jefferson, L.~Martin, H.~McDonald, R.~Morlan, H.~S. Jr., S.~Webb,
  L.~Werdelin, and M.~Wilson (1996).
\newblock Spatial response of mammals to late {Q}uaternary environmental
  fluctuations.
\newblock {\em Science\/}~{\em 272}, 1601--1606.

\bibitem[\protect\citeauthoryear{Green}{Green}{1995}]{green95}
Green, P.~J. (1995).
\newblock Reversible jump {M}arkov chain {M}onte {C}arlo computation and
  {B}ayesian model determination.
\newblock {\em Biometrika\/}~{\em 82}, 711--732.

\bibitem[\protect\citeauthoryear{Hammersley}{Hammersley}{1977}]{hammersley77}
Hammersley, J. (1977).
\newblock Comment on ``spatial contact models for ecological and epidemic
  spread'' by {D}enis {M}ollison.
\newblock {\em J. R. Statist. Soc. B\/}~{\em 39}, 319.

\bibitem[\protect\citeauthoryear{Hazelwood and Steele}{Hazelwood and
  Steele}{2004}]{hazelwood04}
Hazelwood, L. and J.~Steele (2004).
\newblock Spatial dynamics of human dispersals: {C}onstraints on modelling and
  archaeological validation.
\newblock {\em Journal of Archaeological Science\/}~{\em 31}, 669–679.

\bibitem[\protect\citeauthoryear{Holdaway, Fanning, Jones, Shiner, Witter, and
  Nicholls}{Holdaway et~al.}{2002}]{holdaway02}
Holdaway, S., P.~Fanning, M.~Jones, J.~Shiner, D.~Witter, and G.~Nicholls
  (2002).
\newblock Variability in the chronology of late {H}olocene aboriginal
  occupation on the arid margin of {S}outheastern {A}ustralia.
\newblock {\em Journal of Archaeological Science\/}~{\em 29}, 351--363.

\bibitem[\protect\citeauthoryear{Holdaway, Fanning, and Rhodes}{Holdaway
  et~al.}{2008}]{holdaway08}
Holdaway, S., P.~Fanning, and E.~Rhodes (2008).
\newblock Challenging intensification: human-environment interactions in the
  {H}olocene.
\newblock {\em The Holocene\/}~{\em 18}, 403--412.

\bibitem[\protect\citeauthoryear{Hughen, Baillie, Bard, Bayliss, Beck,
  Bertrand, Blackwell, Buck, Burr, Cutler, Damon, Edwards, Fairbanks,
  Friedrich, Guilderson, Kromer, McCormac, Manning, Ramsey, Reimer, Reimer,
  Remmele, Southon, Stuiver, Talamo, Taylor, van~der Plicht, and
  Weyhenmeyer}{Hughen et~al.}{2004}]{hughen04}
Hughen, K., M.~Baillie, E.~Bard, A.~Bayliss, J.~Beck, C.~Bertrand,
  P.~Blackwell, C.~Buck, G.~Burr, K.~Cutler, P.~Damon, R.~Edwards,
  R.~Fairbanks, M.~Friedrich, T.~Guilderson, B.~Kromer, F.~McCormac,
  S.~Manning, C.~B. Ramsey, P.~Reimer, R.~Reimer, S.~Remmele, J.~Southon,
  M.~Stuiver, S.~Talamo, F.~Taylor, J.~van~der Plicht, and C.~Weyhenmeyer
  (2004).
\newblock {Marine04} marine radiocarbon age calibration, 26 - 0 ka {BP}.
\newblock {\em Radiocarbon\/}~{\em 46}, 1059--1086.

\bibitem[\protect\citeauthoryear{Ib{\'{a}\~{n}}ez and
  Sim\'{o}}{Ib{\'{a}\~{n}}ez and Sim\'{o}}{2007}]{ibanez07}
Ib{\'{a}\~{n}}ez, M.~V. and A.~Sim\'{o} (2007).
\newblock A geostatistical spatiotemporal modelling with change points.
\newblock Fifth workshop on Bayesian inference in stochastic processes,
  Valencia 

\bibitem[\protect\citeauthoryear{Jones and Nicholls}{Jones and
  Nicholls}{2001}]{jones01}
Jones, M. and G.~Nicholls (2001).
\newblock Reservoir offset models for radiocarbon calibration.
\newblock {\em Radiocarbon\/}~{\em 43}, 119--124.

\bibitem[\protect\citeauthoryear{Jones and Nicholls}{Jones and
  Nicholls}{2002}]{jones02}
Jones, M. and G.~Nicholls (2002).
\newblock New radiocarbon calibration software.
\newblock {\em Radiocarbon\/}~{\em 44}, 663--674.

\bibitem[\protect\citeauthoryear{Karlsberg}{Karlsberg}{2006}]{karlsberg06}
Karlsberg, A. (2006).
\newblock {\em Statistical modelling for robust and flexible chronology
  building}.
\newblock Ph.d. thesis, University of Sheffield.

\bibitem[\protect\citeauthoryear{Lee}{Lee}{1999}]{lee99}
Lee, T. (1999).
\newblock A stochastic tessellation for modelling and simulating colour
  aluminium grain images.
\newblock {\em Journal of Microscopy\/}~{\em 193}, 109--126.

\bibitem[\protect\citeauthoryear{Majumdar, Gelfand, and Banerjee}{Majumdar
  et~al.}{2005}]{majumdara05}
Majumdar, A., A.~Gelfand, and S.~Banerjee (2005).
\newblock Spatio-temporal change-point modeling.
\newblock {\em Journal of Statistical Planning and Inference\/}~{\em 130},
  149--166.

\bibitem[\protect\citeauthoryear{McCormac, Hogg, Blackwell, Buck, Higham, and
  Reimer}{McCormac et~al.}{2004}]{mccormac04}
McCormac, F., A.~Hogg, P.~Blackwell, C.~Buck, T.~Higham, and P.~Reimer (2004).
\newblock {SHCal04} {S}outhern {H}emisphere calibration 0 - 11.0 cal kyr bp.
\newblock {\em Radiocarbon\/}~{\em 46}, 1087--1092.

\bibitem[\protect\citeauthoryear{M{\o}ller}{M{\o}ller}{1999}]{moller99}
M{\o}ller, J. (1999).
\newblock Topics in {V}oronoi and {J}ohnson-{M}ehl tessellations.
\newblock In W.~K. O.E. Barndorff-Nielsen and M.~van Lieshout (Eds.), {\em
  Stochastic Geometry: Likelihood and Computations}, Monographs on Statistics
  and Applied Probability. Chapman and Hall/CRC.

\bibitem[\protect\citeauthoryear{M{\o}ller and Diaz-Avalos}{M{\o}ller and
  et~al.}{2008}]{moller08}
M{\o}ller, J. and C.~Diaz-Avalos (2008).
\newblock Structured spatio-temporal shot-noise {C}ox point process models,
  with a view to modelling forest fires.
\newblock {\em Scand. J. Statist.\/}.
\newblock to appear.

\bibitem[\protect\citeauthoryear{Mosimann}{Mosimann}{1962}]{mosimann62}
Mosimann, J. (1962).
\newblock On the compound multinomial distribution, the multivariate
  $\beta$-distribution, and correlations among proportions.
\newblock {\em Biometrika\/}~{\em 49}, 65--82.

\bibitem[\protect\citeauthoryear{Naylor and Smith}{Naylor and
  Smith}{1988}]{naylor88}
Naylor, J.~C. and A.~F.~M. Smith (1988).
\newblock An archaelogical inference problem.
\newblock {\em J. Am. Statist. Ass.\/}~{\em 83}, 588--595.

\bibitem[\protect\citeauthoryear{Nicholls and Jones}{Nicholls and
  Jones}{2001}]{nicholls01}
Nicholls, G. and M.~Jones (2001).
\newblock Radiocarbon dating with temporal order constraints.
\newblock {\em Appl. Statist.\/}~{\em 50}, 503--521.

\bibitem[\protect\citeauthoryear{Nunn}{Nunn}{2007}]{nunn07}
Nunn, P. (2007).
\newblock Echoes from a distance: progress report on research into the {L}apita
  occupation of the {R}ove {P}eninsula, southwest {V}iti {L}evu {I}sland,
  {F}iji.
\newblock In S.~C. Bedford, S. and S.~Connaughton (Eds.), {\em Oceanic
  {E}xplorations: {L}apita and {W}estern {P}acific {S}ettlement.}, Volume~26 of
  {\em Terra Australis}, pp.\  163--176. Canberra: Australian National
  University.

\bibitem[\protect\citeauthoryear{Nunn}{Nunn}{2009}]{nunn09}
Nunn, P. (2009).
\newblock Geographical influences on settlement-location choices by initial
  colonizers: a case study of the {F}iji {I}slands.
\newblock {\em Geographical Research\/}~{\em 47}, 306--319.

\bibitem[\protect\citeauthoryear{Nunn, Kumar, Matararaba, Ishimura, Seeto,
  Rayawa, Kuruyawa, Nasila, Oloni, Rati~Ram, Saunivalu, Singh, and Tegu}{Nunn
  et~al.}{2004}]{nunn04}
Nunn, P., R.~Kumar, S.~Matararaba, T.~Ishimura, J.~Seeto, S.~Rayawa,
  S.~Kuruyawa, A.~Nasila, B.~Oloni, A.~Rati~Ram, P.~Saunivalu, P.~Singh, and
  E.~Tegu (2004).
\newblock Early {L}apita settlement site at {B}ourewa, southwest {V}iti {L}evu
  {I}sland, {F}iji.
\newblock {\em Archaeology in Oceania\/}~{\em 39}, 139--143.

\bibitem[\protect\citeauthoryear{Ramsey}{Ramsey}{2001}]{ramsey01}
Ramsey, C.~B. (2001).
\newblock Development of the radiocarbon program {OxCal}.
\newblock {\em Radiocarbon\/}~{\em 43}, 355--363.

\bibitem[\protect\citeauthoryear{Richardson}{Richardson}{1973}]{Richardson73}
Richardson, D. (1973).
\newblock Random growth on a tessalation.
\newblock {\em Proc. Cambridge Philos. Soc.\/}~{\em 74}, 515--528.

\bibitem[\protect\citeauthoryear{Stuiver and Polach}{Stuiver and
  Polach}{1977}]{stuvier77}
Stuiver, M. and H.~Polach (1977).
\newblock Discussion: reporting of 14{C} data.
\newblock {\em Radiocarbon\/}~{\em 19}, 355--363.

\bibitem[\protect\citeauthoryear{Zeidler, Buck, and Litton}{Zeidler
  et~al.}{1998}]{zeidler98}
Zeidler, J., C.~Buck, and C.~Litton (1998).
\newblock The integration of archaeological phase information and radiocarbon
  results from the {J}ama {R}iver {V}alley, {E}cuador: a {B}ayesian approach.
\newblock {\em Latin American Antiquity\/}~{\em 9}, 135--159.

\bibitem[\protect\citeauthoryear{Zhu, Rasmussen, M{\o}ller, Aukema, and
  Raffa}{Zhu et~al.}{2008}]{zhu08}
Zhu, J., J.~Rasmussen, J.~M{\o}ller, B.~Aukema, and K.~Raffa (2008).
\newblock Spatial-temporal modeling of forest gabs generated by colonization
  from below- and above-ground bark beetle species.
\newblock {\em J. Am. Statist. Ass.\/}~{\em 103}, 162--177.

\end{thebibliography}

\end{document}